\documentclass[twocolumn]{article}
\usepackage[a4paper, margin=1in]{geometry}
\usepackage[utf8]{inputenc}
\usepackage{amsmath}
\usepackage{graphicx}
\usepackage{multirow}
\usepackage{float}
\usepackage{cite}
\usepackage{xcolor}
\usepackage{algorithm}
\usepackage{algorithmic}
\graphicspath{ }
\usepackage{comment}
\usepackage{placeins}
\usepackage{hyperref}
\usepackage{fancyhdr}
\hypersetup{
    colorlinks=true,
    linkcolor=blue,
    filecolor=magenta,
    urlcolor=cyan,
    }

\title{Sparse learned kernels for interpretable and efficient medical time series processing}
\author{Sully F. Chen\thanks{Duke University School of Medicine, Durham, NC 27708, USA. (corresponding author, e-mail: sully.chen@duke.edu)} 
\and Zhicheng Guo\thanks{Department of Electrical and Computer Engineering, Duke University, Durham, NC 27708, USA. (e-mail: zhicheng.guo@duke.edu)} 
\and Cheng Ding\thanks{Department of Biomedical Engineering, Georgia Institute of Technology \& Emory University, Atlanta, GA 30322, USA. (e-mail: chengding@gatech.edu)} 
\and Xiao Hu\thanks{Nell Hodgson Woodruff School of Nursing, Emory University, 1520 Clifton Rd, Atlanta, GA 30322, USA. (e-mail: xiao.hu@emory.edu)} 
\and Cynthia Rudin\thanks{Department of Computer Science and Department of Electrical and Computer Engineering, Duke University, Durham, NC 27708, USA. (corresponding author, e-mail: cynthia@cs.duke.edu)}}

\date{}

\begin{document}

\maketitle

\thispagestyle{fancy}

\begin{abstract}
Rapid, reliable, and accurate interpretation of medical time-series signals is crucial for high-stakes clinical decision-making. Deep learning methods offered unprecedented performance in medical signal processing but at a cost: they were compute-intensive and lacked interpretability. We propose \textit{Sparse Mixture of Learned Kernels} (SMoLK), an interpretable architecture for medical time series processing. SMoLK learns a set of lightweight flexible kernels that form a single-layer sparse neural network, providing not only interpretability, but also efficiency, robustness, and  generalization to unseen data distributions. We introduce a parameter reduction techniques to reduce the size of SMoLK's networks while maintaining performance. We test SMoLK on two important tasks common to many consumer wearables: photoplethysmography (PPG) artifact detection and atrial fibrillation detection from single-lead electrocardiograms (ECGs). We find that SMoLK matches the performance of models orders of magnitude larger. It is particularly suited for real-time applications using low-power devices, and its interpretability benefits high-stakes situations.
\end{abstract}

\section{Introduction}
Medical signals such as electroencephalograms (EEGs), electrocardiograms (ECGs), and photoplethysmography signals (PPGs) have played a pivotal role in the field of medicine, serving as a crucial medium for diagnosing and monitoring patients' conditions, facilitating timely interventions, and improving overall healthcare outcomes. These types of signals are usually manually reviewed by trained medical professionals, however, in today's realm of endless medical data and the popularization of continuous passive monitoring, manual reviewing efforts have become inefficient or even a bottleneck in the patient care process. Traditional techniques to automate signal processing and categorization, while improving speed, often come with limitations in terms of performance. The rise of deep learning yielded a revolution in signal processing, offering models with unparalleled performance (e.g., \cite{egger, nogales, shin, pereira2019deep, liu2020}).
However, these models often come with their own set of challenges, particularly in terms of computational demands and a lack of interpretability in their decision-making processes. Post-hoc saliency methods often fail to correctly identify feature importance for time-series signals from these black box models \cite{ismail2020, turbe2023}.

Recognizing the need for high-performing interpretable models, we sought to design an architecture that combines the best of both worlds: the performance of deep learning with the computational efficiency and interpretability of traditional methods. 
We propose a method that has performance similar to -- and often better than -- state-of-the-art deep neural network (DNN) approaches but with several orders of magnitude fewer parameters. Our method, \textit{Sparse Mixture of Learned Kernels (SMoLK)} learns a set of lightweight flexible kernels to construct a \textit{single-layer sparse} neural network, providing a new efficient, robust, and interpretable medical signal processing architecture. To train these kernel functions, we introduce two parameter reduction techniques: \textit{weight absorption} absorbs the kernel weighting factors into the kernels themselves and \textit{correlated kernel pruning} reduces redundancies among the learned kernels.

While our architecture is versatile enough for a range of applications, we chose to spotlight its efficacy using photoplethysmography (PPG) and single-lead electrocardiography (ECG), given their increasingly widespread integration into wearable devices and use for long-term cardiac monitoring. Photoplethysmography poses many challenges, such as motion artifacts and signal noise, which are commonly encountered in various other medical signals as well. Photoplethysmography is a non-invasive optical technique to measure blood volume changes in tissue by measuring changes in light absorption. It is commonly used to infer various cardiovascular parameters, such as blood oxygenation, heart rate, heart rate variability, and other related parameters \cite{ray2021, hoogantink2021, ding2021, spierer2015, nabeel2017, nilsson2013, castaneda}. It has become abundant in wearable consumer devices (i.e., smart watches) and there is a continuing effort to develop algorithms that extract meaningful data from PPG signals obtained from these devices.

A notable limitation of wearable PPG devices is their sensitivity to motion-induced artifacts, which leads to corruption in the collected signals. Motion artifacts occur commonly since most users are not stationary. Motion artifacts often harm measurement quality and it requires considerable effort to identify and either discard or reconstruct these corrupted segments of the time series. Previous methods relied on either additional sensors (e.g., accelerometers) \cite{accel_PPG},  statistical methods such as standard deviation, skew, kurtosis, wavelet-based motion artifact reduction algorithms \cite{stat_1, stat_2} or more sophisticated handcrafted feature detectors \cite{handcrafted_1, handcrafted_2}. The aforementioned hand-crafted-feature-based-detectors and statistical methods provide clear indications and reasoning for why a segment is identified as an artifact, but often underperform compared to other methods, such as deep neural networks \cite{esgalhado, pankaj, ismail}. However, deep neural nets are computationally intensive and contain millions of parameters, thus making them incompatible with small, low-power wearable devices. Furthermore, deep neural networks lack interpretability; it is difficult, if not impossible, to determine a rationale behind the identification of an artifact in the PPG signal. For the same reason, they are difficult to troubleshoot.

Similar to PPG, ECG is also a vital diagnostic and monitoring tool for cardiac events, especially in hospital settings. By measuring the timing and rhythm of each electrical impulse from the heart, it provides crucial information about the heart's condition, helping to detect various cardiac abnormalities such as arrhythmias, heart attacks, and other heart diseases. ECG is considered to be the gold standard for diagnosing a variety of cardiac events. There have been several works in automating disease detection and diagnosis from ECG signals using deep learning models \cite{ecgdl1, ecgdl2, ecgdl3}; however, these methods share similar issues to PPG-based deep nerual networks mentioned above -- they are computationally inefficient and lack interpretability.

Our approach achieves compelling results in both the task of detecting artifacts in PPG signals, as well as the detection of atrial fibrillation from single-lead ECGs (an increasingly common task in modern-day wearables). For PPG artifact segmentation, SMoLK obtains the best of all worlds: (1) \textit{state-of-the-art (SotA) performance} and (2) \textit{interpretability} in a model with (3) \textit{several orders of magnitude fewer parameters} than the current SotA. For the single-lead ECG task, SMoLK achieves metrics matching a deep ResNet with less than 1\% the parameter count, while significantly outperforming deep networks in the low-data regime, even against equal-parameter deep networks controlling, for potential overfitting. For interpretability (2), SMoLK yields a tiny-yet-optimized set of convolution kernels that, when applied to the PPG signal, summed, and ceiled, directly produces a measure of PPG signal quality. This measure can then be thresholded to segment artifacts. The process is ``nearly linear'' given that the only non-linearity is a ceiling function setting negative values to zero. We can \textit{directly and precisely measure the contribution of any individual convolution to the predicted signal quality}, and kernel contributions are directly proportional to the output signal. For classification tasks, we can directly invert the process to compute each part of the signal's precise contribution to the final classification. Our approach is reminiscent of feature-detection-based approaches in that we learn a set of kernels that mimic handcrafted features. For our smallest model, we learn so few kernels (12 total) that we can inspect these features by eye and observe the waveforms learned by our approach. This contrasts with deep neural networks, where there are so many parameters one cannot feasibly inspect the inner workings of the model, and even if one took the time to meticulously inspect these features, it would still be unclear how each feature individually contributes to the output due to the built-in non-linearity.

In terms of efficiency and memory (3), our model reflects that for real-world applications, signal processing can be limited by computational resources or power consumption, as most wearable devices are small computing devices with limited battery life and low compute budget. Furthermore, memory is often a constraint as many wearable devices run multiple processes and monitor several metrics at once.
SMoLK is simple to implement, and easily lend itself to memory and compute optimization.  
Our architecture can be quantized easily to 16-bit floating point operations without any appreciable decrease in performance. 
Our method is lightweight enough to allow real-time assessment of signal quality or classification of arrhythmias with a small enough impact that it could feasibly be implemented as a background process, using only kilobytes of memory.

\section{Related Work}

\subsection{Classical Statistical and Machine Learning Techniques}
Statistical and machine learning techniques involve applying various algorithms to extract features and build classifiers for detecting motion artifacts in PPG signals. Several studies have used least-squares-based methods, such as X-LMS \cite{tanweer} or adaptive filters \cite{wu}. Other methods have used statistical parameters like kurtosis or Shannon entropy as features to detect artifacts \cite{selvaraj, hanyu}. Additionally, machine learning techniques, such as support vector machines, random forests, and na\"ive Bayes have been employed \cite{lin, athaya} for this task. For ECG arrhythmia detection, several works also leveraged support vector machines, wavelet transforms and logistic regression \cite{svmecg, waveletecg, bazi2013domain}.
These methods often under-perform compared to deep learning-based approaches and typically focus on the classification of discrete time-chunks rather than segmentation of artifacts.

\subsection{Deep Learning Approaches}
Methods such as 1D convolutional neural networks \cite{goh, yalizheng, ecg1dcnn, ecg1dcnn2}, 2D convolutional neural networks \cite{liu, ecg2dCNN}, and U-Net type architectures \cite{sota} have been implemented with varying degrees of success for PPG artifact detection and ECG arrhythmia detection. These methods are more accurate than traditional machine learning methods. Many of these architectures, such as 1D and 2D convolutional neural networks are not inherently designed for segmentation tasks, as they were originally designed for classification tasks. As a result, they must be adapted to segmentation tasks via classification of signal windows, or methods like GradCAM \cite{gradcam} or SHAP \cite{shap}, which adds another layer of complexity. Additionally, these methods are often unreliable (see experiments in \cite{sota}). Transformer-based approaches have been implemented, performing similarly to convolutional networks (though with higher parameter counts) \cite{che2021}.

Most deep learning methods demand significant computational resources, rendering them impractical for use in wearable devices. An exception is Tiny-PPG \cite{yalizheng}, which emphasizes model pruning and the development of compact, portable models. Our experiments indicate that we achieve similar performance to Tiny-PPG but with much sparser models (about half the parameters). However, Tiny-PPG is not designed for interpretability. Trying to understand black box deep learning models is extremely challenging. Most work turns to methods that seek to explain black box model behavior, such as the aforementioned GradCAM and SHAP methods \cite{gradcam, shap}, rather than build interpretability inherently into the model, which yields a faithful understanding of model behavior. Indeed, even modern explanation algorithms for medical signal processing with transformer architectures \cite{vaid2023} fail to yield saliency maps that properly attribute features to their classification. In the aforementioned work, saliency maps for ST-elevation myocardial infarction correctly assign high saliency to \textit{some} elevated ST segments, yet ignore equally prevalent ST segments seemingly at random. Furthermore, an area of equally intense saliency is assigned to a completely blank area of the ECG strip. Although it is not always obvious a priori what features are important, it is generally useful to know what features contributed to a model's decision.

\section{Model Architecture}
\subsection{Overview}
In this study, we framed the task of PPG quality classification as a feature-detection task. We hypothesized that clean PPG signals could be identified by certain common features, and PPG signals with artifacts could be identified by either the lack of these recurring features or by features caused by factors such as movement. Our PPG signals are one-dimensional time-series signals, and we learned a set of kernels to identify these features.

\begin{figure*}[h]
\centering
\includegraphics[width=16cm]{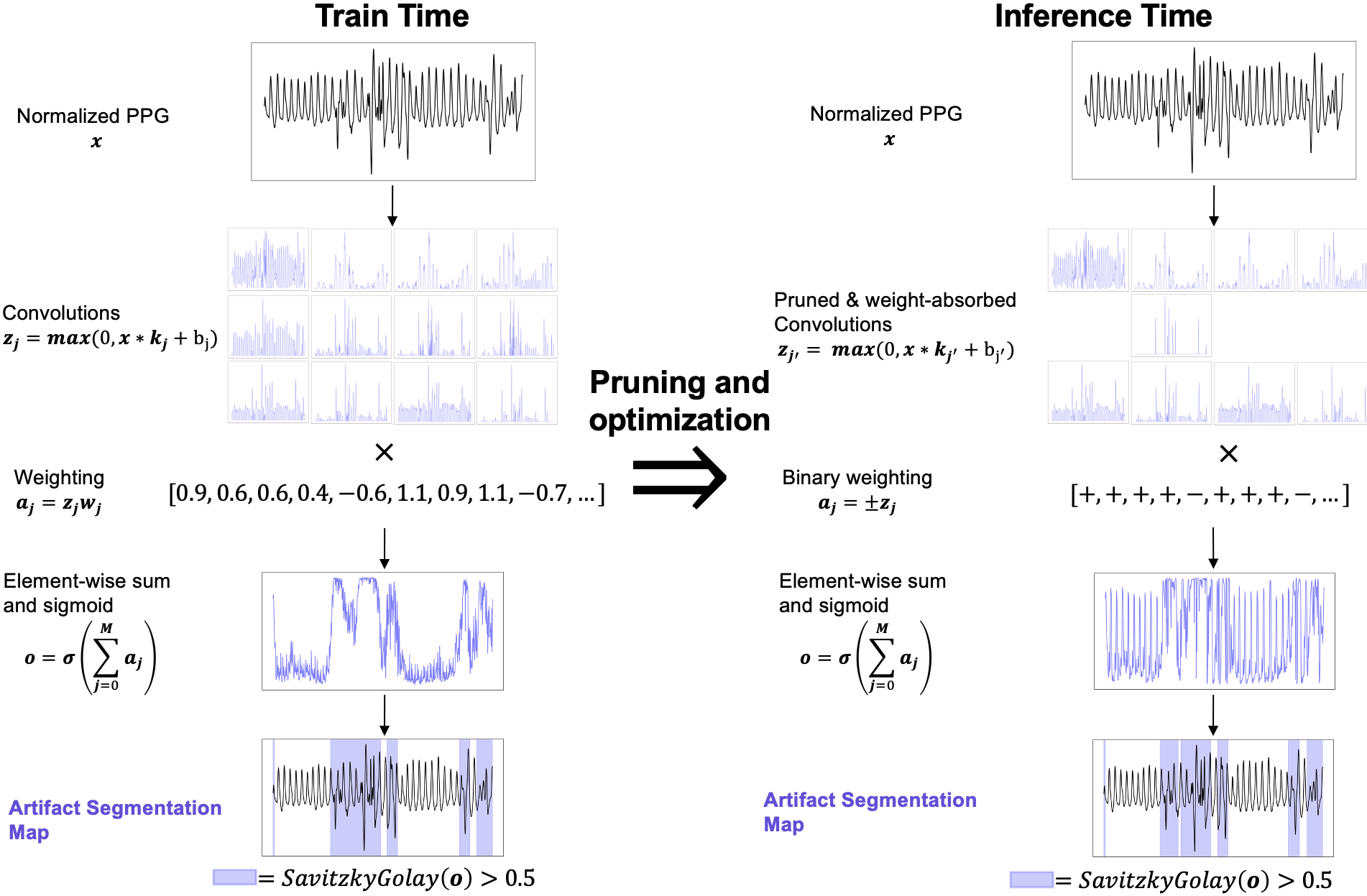}
\caption{\textbf{The PPG Processing Pipeline.} The processing pipeline for the PPG signal quality segmentation task during training and during inference. First, a PPG signal is normalized to a unit normal by subtracting the mean and dividing by the standard deviation. Next, a set of convolutions is applied. Our smallest model is lightweight enough that \emph{all of the convolved signals} are displayed in this figure. The convolved signals are ceiled, weighted, and summed, and a sigmoid is applied. Finally, the output is smoothed and thresholded to segment the signal. After training, similar kernels are ablated and kernel weights are absorbed to reduce parameter count.
\label{fig:pipeline}}
\end{figure*}

To achieve this, we learn a set of $M$ convolution kernels of varying numbers to evaluate parameter scaling laws: $M = 12$ (small), $72$ (medium), $384$ (large). The $M$ kernels are divided into sets of $M/3$ kernels of short (1.0 seconds), moderate (1.5 seconds), and long (3.0 seconds) sizes, with the goal of learning features of different lengths to improve classification quality. The kernels are convolved with the input PPG signal and the values are ceiled via $\max(0, x)$ to obtain a non-negative signal quality indicator. Additionally, a bias is added to each convolution channel. We compute a weighted sum of the convolutions to produce a single channel that represents a measure of signal quality, which is then transformed to range [0.0, 1.0] via the sigmoid function. We apply a 3rd-order Savitzky-Golay filter ($\sim$0.8-second filter window) to smooth the output, then apply thresholding at a value of 0.5 to produce a binary output segmentation. This pipeline is shown in Figure \ref{fig:pipeline}. The addition of a post-processing smoothing filter (Savitzky-Golay in our case) yields smoother segmentation maps and relies on the assumption that PPG artifacts tend to occur in groups, rather than as fragmented artifacts. However, this post-processing filter is not strictly necessary to maintain good DICE metrics and is an optional hyperparameter to be chosen for the use case (see Appendix \ref{sec:appdx_postprocessing}).

More formally, our model consists of $M$ kernels, $M$ scalar biases, and $M$ scalar weights. Given an input signal, $\mathbf{x}$, our SMoLK model is defined as follows:
\begin{align*}
  \text{SMoLK}(\mathbf{x}) &= \sigma\left(\sum^M_{m=0} \text{max}(0, \mathbf{x}*\mathbf{k}_m + b_m) \cdot w_m \right)
\end{align*}
where $\mathbf{k_\textit{m}}$, $b_m$, $w_m$ are the $m^{th}$ convolution kernel, bias, and signal weight respectively, $*$ represents the convolution operation, $\cdot$ is scalar-vector multiplication, and $\sigma$ is the element-wise sigmoid function, $(1+e^{-\mathbf{x}})^{-1}$. 
In essence, SMoLK is equivalent to a linear combination of filtered signals generated by convolving the input with a learnable kernel, mapped to range $(0, 1)$. 
For the segmentation task, the output can be further post-processed via smoothing and thresholding to yield a segmentation map.

\subsection{Classification}
\subsubsection{Overview}
\begin{figure*}[h]
\centering
\includegraphics[width=16cm]{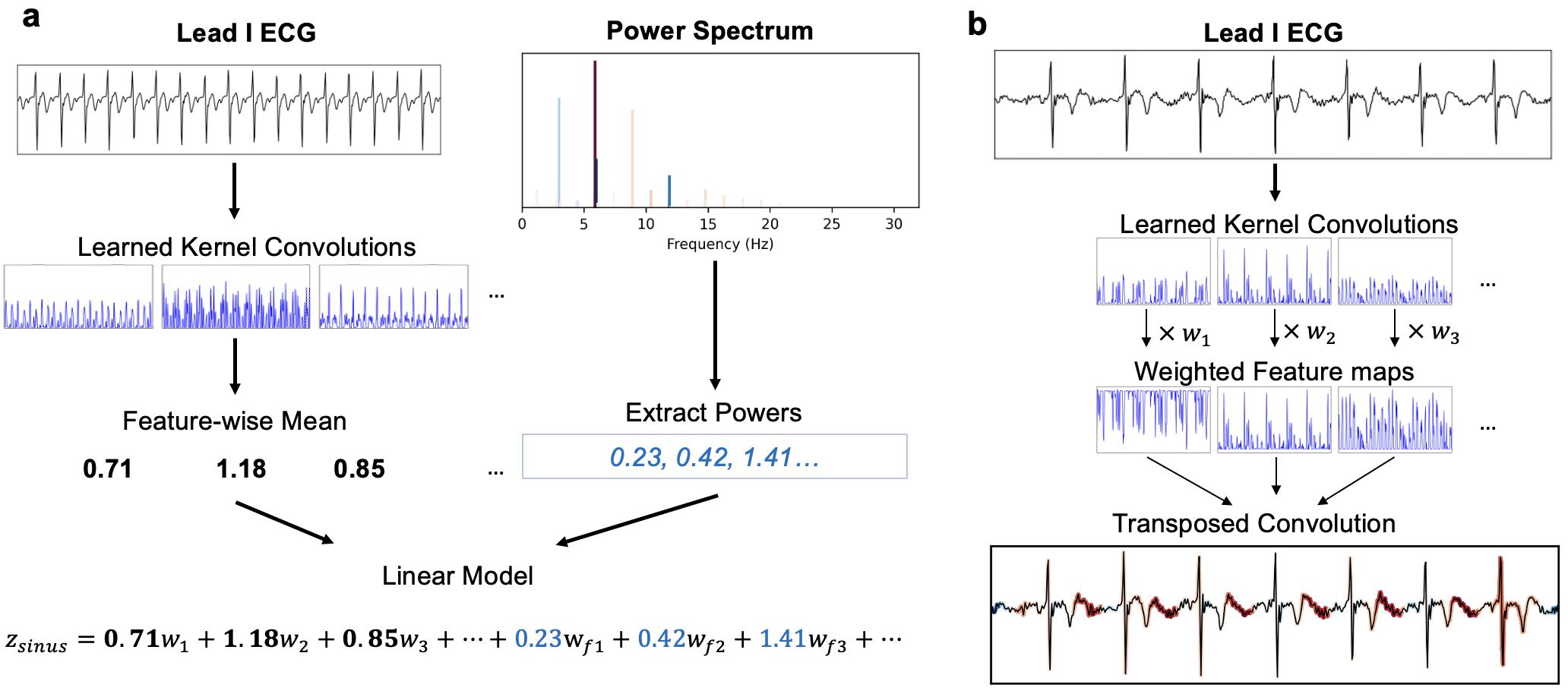}
\caption{\textbf{The ECG Processing Pipeline.} \textbf{a.} The pipeline for processing a single lead ECG. First, a set of learned convolutions is applied to the ECG to produce several feature maps. Then, feature-wise means are computed, which are used as inputs to a linear model. Global frequency information, obtained from a power spectrum, is also used as inputs for the linear model to yield a logit. \textbf{b}. The reverse pipeline for interpreting the prediction in subfigure \textbf{a} from the learned kernel model. First, feature maps are generated by the learned kernel convolutions, in the same manner as \textbf{a}. However, instead of computing the mean value of this feature map, the feature maps are multiplied element-wise with their corresponding class weights. The power spectrum is not used in the interpretation. Lastly, a transposed convolution is applied to compute the importance of each part of the input signal. This method is almost an exact inverse of the forward procedure, allowing for a principled assignment of importance to the input signal that directly correlates to the output class logit. Notably, this method correctly labels the increased PR-interval as the most important feature in the classification of this ECG as $1^{st}$-degree heart block, indicating that our model has learned important features rather than spurious correlations.
\label{fig:ecg_pipeline}}
\end{figure*}

To demonstrate the versatility of our architecture, we slightly modify our model's output to suit a classification task. First, we generate feature maps with our learned kernels:
\begin{align*}
  \mathbf{f_m}(\mathbf{x}) &= \text{max}(0, \mathbf{x}*\mathbf{k}_m + b_m).
\end{align*}
However, instead of computing a weighted element-wise sum of the feature maps (as is performed in the PPG task), we compute the mean value of each individual feature map to produce a score for each feature map:
\begin{align*}
  \phi_m(\mathbf{x}) &= \text{Mean}(\mathbf{f_m}(\mathbf{x}))
\end{align*}
where $\text{Mean}$ computes the element-wise mean of the feature map, $\mathbf{f_m}$. We then use these scores in a simple linear model to produce a logit:
\begin{align*}
  z_j(\mathbf{x}) &= \sum_{m=0}^{M}w_{mj}\phi_m(\mathbf{x}) + b_j
\end{align*}
where $w_{mj}$ is the weight of the $m^{th}$  feature for the $j^{th}$ class, and $b_j$ is the bias for this class.  
In the case of ECG classification, we add global frequency information to our linear model by computing the power spectrum of the ECG signal and adding these as linear components:
\begin{align*}
  z_j(\mathbf{x}) &= \sum_{m=0}^{M}w_{mj}\phi_m(\mathbf{x})  + \sum_{\hat{f} \in \mathcal{F}} w_{\hat{f}j}\hat{f}(\mathbf{x})  + b_j
\end{align*}
Where $\hat{f}(\mathbf{x}) $ are the frequency bands of $\mathbf{x}$ in the power spectrum $\mathcal{F}$, and $w_{\hat{f}j}$ is the corresponding weight of that frequency band for class $j$. Finally, a non-linear activation function can be applied to this logit to generate a probability distribution. We use SoftMax in our case, which is defined as \(\text{SoftMax}(z_j) = \frac{e^{z_j}}{\sum_{{j'}=1}^n e^{z_{j'}}}\), where \(z_j\) is the logit for class $j$. An illustrated overview of this pipeline can be seen in Figure \ref{fig:ecg_pipeline}a.

\subsubsection{Interpretability}
Because our model is a linear combination of feature map mean activations, and each activation was generated from a convolution over the signal, we can directly compute the contribution of each part of the input signal to a given output class by reversing the convolution. We thus compute the contribution map as follows, in Algorithm \ref{alg:contribution}. At a high level, a forward pass is performed as usual and the feature map values are computed. Then, the feature maps are multiplied element-wise with the feature weight for the class in question. Lastly, this product is convolved backward into a contribution map buffer of equal dimension to the input ECG. This allows us to compute exactly how much any given portion of the input signal contributed to the output classification. An illustration of this process can be seen in Figure \ref{fig:ecg_pipeline}b.

\begin{algorithm}
\caption{Compute contribution map for ECG trace}
\begin{algorithmic}[1]
\REQUIRE 
    \STATE ECG trace $\mathbf{x}$ of length $L$.
    \STATE $M$ feature maps denoted as $\mathbf{F} = \{\mathbf{f}_1, \mathbf{f}_2, \ldots, \mathbf{f}_M\}$.
    \STATE Class of interest $j$.
    \STATE Contribution weights $w_{mj}$ denoting the weight of feature map $\mathbf{f}_m$ towards class $j$.
    \STATE Kernel size $K$.
    \STATE Define $\mathrel{+}$ as the element-wise addition of a scalar to a vector.

\STATE Initialize $\mathbf{I}_\mathbf{x}^j$ as a zero vector of size $L$.

\FOR{$m = 1$ to $M$}
    \FOR{$l = 1$ to $L - K + 1$}
        \STATE $\mathbf{I}_\mathbf{x}^j[l : l + K] \mathrel{+}= \mathbf{f}_m[j] \times w_{mj} \times K^{-1}$
    \ENDFOR
\ENDFOR

\RETURN $\mathbf{I}_\mathbf{x}^j$
\end{algorithmic}
\label{alg:contribution}
\end{algorithm}

\subsection{Weight Absorption}
We can reduce the parameter count of our segmentation model by absorbing the weighting factor into the kernels themselves, due to the ``almost linear'' property of the model. Let $\mathbf{x}$ be the input PPG signal, which is a vector of length 1920. Let $\mathbf{k}_m$, $b_m$, $w_m$ be the $m^{th}$ convolution kernel, bias, and signal weight respectively. The kernels are vectors of length 192, 96, or 64 depending on the length class (long, moderate, or short), while the biases and weights are scalars. We have:
\begin{eqnarray*}
\lefteqn{
  SMoLK(\mathbf{x}) }\\
 & \hspace*{-30pt}=& \hspace*{-20pt}\sum^M_{m=0} \text{max}(0, \mathbf{x}*\mathbf{k}_m + b_m) \cdot w_m \\
  &\hspace*{-30pt}=& \hspace*{-20pt}\sum^M_{m=0} \text{max}(0, (\mathbf{x}*\mathbf{k}_m + b_m) \cdot |w_m|) \cdot \text{sgn}(w_m) \\
  &\hspace*{-30pt}=& \hspace*{-20pt}\sum^M_{m=0} \text{max}(0, (\mathbf{x}*\mathbf{k}_m)\cdot|w_m| + b_m \cdot |w_m|) \cdot \text{sgn}(w_m) \\
  &\hspace*{-30pt}=& \hspace*{-20pt}\sum^M_{m=0} \text{max}(0, [\mathbf{x}*(\mathbf{k}_m\cdot|w_m|)] + b_m \cdot |w_m|) \cdot \text{sgn}(w_m),
\end{eqnarray*}
where $|\cdot |$ is  element-wise absolute value.

The quantity $\mathbf{k}_m\cdot|w_m|$ becomes the new kernel, and the quantity $b_m \cdot |w_m|$ becomes the new bias. We can thus reduce parameter count, since we can now store the kernels themselves along with a corresponding binary value indicating whether they are positive or negative kernels. Furthermore, each kernel can be neatly categorized into ``contributing to good signal quality'' (positive kernels) or ``contributing to poor signal quality'' (negative kernels). This process is performed after learning the kernels to reduce the memory footprint and number of scalar operations for deployed models.

\subsection{Correlated Kernel Pruning}
Driven by the observation that some learned kernels look quite similar, we hypothesized that we could prune similar kernels while maintaining good performance. To measure the similarity between kernels, we use Euclidean distance.

At a high level, our kernel pruning method consists of the following steps:
\begin{enumerate}
    \item For each convolutional layer, compute the Euclidean distance between all pairs of kernels.
    \item Sort the kernel pairs by their Euclidean distance (in order of increasing distance), and select the top $pa$ pairs to prune, where $pa$ is a hyperparameter. 
    \item For each selected pair to prune, determine which of the two kernels to remove and which to keep based on their effective contribution to the signal.
\end{enumerate}

A rigorous description of the pruning process is discussed in the Methods section.

\section{Methods}
For all machine learning tasks, PyTorch 2.1.1 \cite{pytorch} and Scikit-learn 1.3.2 \cite{scikit-learn} were used. Data analysis and statistics were conducted with SciPy 1.9.3 \cite{scipy} and NumPy \cite{numpy} 1.22.4.

\subsection{Datasets and Preprocessing}
\label{sec:appdx_datasets_and_preprocessing}
We train our model on the PPG-DaLiA \cite{dalia} training split, and evaluate our approach on the same datasets used by the current state-of-the-art algorithms \cite{sota, yalizheng}, namely PPG-DaLiA \cite{dalia} test split, WESAD \cite{wesad}, and TROIKA \cite{troika}. The PPG-DaLiA dataset is composed of recordings from 15 subjects performing real-world tasks. The data include electrocardiogram (ECG), accelerometer, and electrodermal recordings for various settings/activities, such as sitting still, walking up/down stairs, playing sports (soccer, cycling, walking), and driving. The WESAD data was recorded from wrist and chest PPGs from 15 subjects aged 21-55 (median 28) primarily focusing on varying the emotional state of the subject (neutral, stressed, amused). ECG, accelerometer, electrodermal, and wrist PPG signals were recorded. Lastly, the TROIKA data was recorded from 12 subjects running on a treadmill (ages 18-35). Accelerometer, ECG, and PPG signals were recorded. This dataset represents the most challenging dataset for generalization among the datasets we test in this work, since the subjects were under intense physical demands with poor signal quality, owing to the large amounts of movement.

We followed the same preprocessing procedure as Guo et al. \cite{sota}, namely applying a bandpass filter with cutoffs of 0.9 Hz and 5 Hz, splitting the data into 30-second chunks, and resampling to 64 Hz. However, instead of normalizing the chunks between $[0, 1]$ as Guo et al$.$ did, we instead opted to normalize each chunk to a unit normal distribution. Normalization was performed at the chunk-level rather than the dataset-level to ensure that signal intensities remained relatively consistent across chunks, improving generalization to distributions of PPG signals that may have different global parameters or greater inter-sample variability. We used Guo et al.'s publicly available artifact segmentation labels, which were created via a web-annotation tool and a group of skilled annotators \cite{sota}.

For our experiments with ECGs, we focus on the task of single-lead atrial fibrillation detection, an increasingly common feature of wearable devices. We trained on the Computing in Cardiology dataset \cite{cinc}, which consists of 5154 sinus rhythm recordings, 771 recordings of atrial fibrillation, 2557 recordings of ``other'' arrhythmias (unspecified), and 46 recordings that were too noisy to classify, all recorded at 300 Hz. The recordings vary in length, with the average lengths being 31.9, 31.6, 34.1, and 27.1 seconds for each of the aforementioned classes, respectively. For our analysis, we excluded the ``noisy'' class from our task, as our holdout set does not include a noisy class. Although our methodology can classify ECG segments of arbitrary length (unlike many deep learning-based approaches, which require the input lengths to be fixed), we chose to split the dataset into 10-second chunks to more easily compare our results to deep-learning based methods. After splitting the data into 10-second chunks, we obtained 16742 sinus samples, 2463 atrial fibrillation samples, and 8685 ``other'' samples. We applied a bandpass filter with cutoffs at 1.0 and 10.0 Hz, as well as per-sample normalization to a unit normal (subtracting the mean and dividing by the standard deviation).

As a holdout set to test the generalization of our method, we used a large, publicly available dataset collected from Chapman University, Shaoxing People’s Hospital, and Ningbo First Hospital \cite{Zheng2020, Zheng2022}. This is a diverse 12-lead ECG dataset, and we used arrhythmias classified as either atrial fibrillation (1780 samples), sinus rhythm (7858 samples), or other (11695 samples). Each sample consists of a 10-second, 500 Hz 12-lead recording, of which we took only lead I for our single-lead task. We processed the data identically to the Computing in Cardiology dataset.

We refer to the original works for detailed statistics regarding the datasets \cite{Zheng2020, Zheng2022, cinc}.

\subsection{PPG Segmentation}
The following experiments aim to compare our methodology to several common baselines, as well as a state-of-the-art multi-million parameter architecture based on U-Net. We show that our segmentation algorithm matches or exceeds the state-of-the-art on several benchmark PPG tasks.

\subsubsection{Training}
To learn the kernels, we used stochastic gradient descent with Adam optimization \cite{kingma} ($\beta_1=0.9$, $\beta_2 = 0.999$, $\epsilon=10^{-8}$, weight decay $=10^{-4}$), and a linear learning rate decay schedule (decaying from 0.01 to 0.002 learning rate on the final iteration), with a binary cross-entropy error objective function. Since our dataset is on the order of $10^8$ bytes and our model sizes are on the order of $10^3$--$10^5$ bytes, we can compute the gradient over the entire dataset in a single pass to obtain the true gradient (as opposed to an estimate via a mini-batch). For the largest model sizes, we use gradient accumulation due to memory constraints. Thus, we train for 512 iterations computing the gradient over the entire training set. We find that this yields empirically better performance than stochastic mini-batches. We find it empirically difficult to overfit the training data given our parameter count no matter the iteration count, especially at smaller kernel numbers.

\subsubsection{Pruning}
To determine which kernel to prune, we compute the ``effective contribution'' of the kernel as:
\begin{equation}
\text{eff}_j = w_j \cdot \mu_j, \quad \text{eff}_k = w_k \cdot \mu_k
\end{equation}
where $w_j$ and $w_k$ are the weights of the two kernels in a pair and $\mu_j$ and $\mu_k$ are the mean absolute values of the kernels $w_j$ and $w_k$, respectively. We keep the kernel with the higher effective weight and prune the other kernel.

When pruning a kernel, we need to update the weight and biases of the remaining kernel to correct for the removed kernel's contribution. Without loss of generality, assume that $\text{eff}_j > \text{eff}_k$. We update the weight and bias of the remaining kernel $j$ as follows:
\begin{equation}
w_j' = \frac{\text{eff}_j + \text{eff}_k}{\mu_j}, \quad b_j' = b_j + b_k.
\end{equation}
where $b_j$ and $b_k$ are the biases for weight $w_j$ and $w_k$ respectively. In essence, the bias of the two kernels are simply combined into one term to compensate for pruning one kernel, and the weight of the remaining kernel is increased proportionally to account for the loss in signal from the pruned kernel. This technique allows for the removal of 22\% of the parameters of our largest model, with less than a 2\% absolute impact on performance on all datasets (Appendix Table \ref{tab:appdx_pruning_metrics}). Moreover, this technique is adaptable, allowing for the removal of more parameters at a greater performance cost, or a smaller number of parameters at a lesser performance cost. Notably, the choice of Euclidean distance is arbitrary, and any distance metric can be chosen for the pruning step. Indeed, one can use a variety of distance metrics with similar results.

We provide three example pruning results at various model scales, with a target performance decrease of no greater than 5\% (Appendix Table \ref{tab:pruning}). We would like to note that one can experiment with this pruning process via our open-source GitHub repository.

Furthermore, we show pruning results using various distance metrics perform relatively consistently, indicating a robustness to the chosen distance metric. We test three additional distance metrics on our large model: cosine similarity between power spectra (cosine distance in frequency space), cosine similarity in feature space (the cosine similarity of the feature vectors themselves), and Manhattan distance (Appendix Table \ref{tab:appdx_pruning_metrics}).

\subsubsection{Evaluation}
To evaluate the performance of the baselines and learned kernels for artifact detection, we used the DICE score, defined as
\[
DICE(A, B) = \frac{2|A \cap B|}{|A| + |B|}
\]
where \(A\) and \(B\) are the binary segmentation maps of the predicted and ground truth, respectively, and \(|\cdot|\) denotes the cardinality of a set. We employed this score on a test set of PPG signals, with each 30-second chunk of PPG signal normalized via the method described earlier. We conducted 10-fold cross-validation using 10 different models trained from different starting seeds.

\subsubsection{Baselines}
We compare to the same baselines tested in Guo et al. \cite{sota}. Namely, a convolutional neural network sliding window approach, a template-matching approach, and a ResNet-34-based classifier with segmentation performed via GradCAM \cite{gradcam} or SHAP \cite{shap}. Further details are given in the Baseline Methods section.

\subsection{Atrial Fibrillation Detection}
\subsubsection{Training}
We train our learned-kernel atrial fibrillation detector similarly,  with AdamW optimization \cite{adamw} ($\beta_1=0.9$, $\beta_2 = 0.999$, $\epsilon=10^{-8}$, weight decay $=10^{-2}$), and a linear learning rate decay schedule (decaying from 0.1 to 0.0 learning rate on the final iteration) for 512 epochs, with a multi-class cross-entropy error objective function. However, due to memory constraints, we train with a batch size of 1024 rather than over the entire dataset.

\subsubsection{Evaluation}
We compute F1-score and AUC-ROC as our primary metrics. We train our models on the Computing in Cardiology dataset, optimizing for 10-fold cross-validation F1-scores. We then test our models on an entirely unrelated test set constructed from the Zheng et al$.$ dataset previously mentioned \cite{Zheng2020, Zheng2022}.

\subsubsection{Baselines}
We compare against a common deep learning architecture for medical signal processing, namely a 1D-ResNet, as well as several previously state-of-the-art low-parameter count convolutional neural networks. These baselines are outlined in the Baseline Methods section.

\subsubsection{Interpretability}
We trained another learned kernel classifier to distinguish between $1^\circ$ atrioventricular block (1140  samples) and sinus rhythm (8125  samples) to demonstrate the interpretability of our method. We chose  $1^\circ$ atrioventricular block to demonstrate the interpretability of our method as it is defined by a single salient feature (the PR-interval), rather than a set of global features (e.g., irregularly irregular beats in atrial fibrillation). We gather this data from the Zheng et al$.$ dataset.

\subsection{Baseline Methods}
\subsubsection{Convolutional Sliding Window}
Our first baseline is a 1D-convolutional network trained to classify 3-second windows of PPG signal as either ``artifact'' or ``clean.'' The network consists simply of 3 blocks of convolution-ReLU-BatchNorm-MaxPool. The 3 blocks consist of kernel sizes 10, 5, and 3, and the channel numbers are 64, 64, and 128. The model was trained from 5000 three-second windows that were randomly selected from the PPG-DaLiA training set. Training consisted of 200 epochs of minimizing cross-entropy loss with Adam. Testing was done by segmenting the 30-second PPG signal into 3 second chunks with the trained convolutional neural network classifier, and a sliding window with a step size of 1 second was used for testing. Since the trained convolutional network has an input size of 3-second chunks, each second of PPG signal receives 3 classification outputs. A 1-second chunk of PPG signal is classified as an artifact if \textit{any} of the three classification outputs are classified as an artifact.

\subsubsection{Template-Matching}
We follow the same approach as Guo et al. \cite{sota}, which was inspired by Lim et al.\cite{lim}. Signals were first divided from the PPG-DaLiA training set into separate pulses via peak detection. 10 clean (artifact-free) pulses were then selected to serve as our standard templates for comparison with test pulses. Each pulse from the test data was compared to these 10 templates, using the dynamic time warping (DTW) distance metric to measure similarity.
The smallest DTW distance (out of the 10 comparisons with the templates) was then identified. The range of the DTW distance function is $[0, \infty)$, but empirically we find that the DTW distances between our templates and the PPG signals fall mostly within the range $[0, 10]$. We created a binary predicted label by labeling a segment as an artifact if the minimum DTW distance is at least 1. Other samples were given ``non-artifact'' predicted labels.
The threshold of DTW was chosen by testing thresholds between 0 and 10 over the entire train set and choosing the threshold $\beta$ that produced the highest DICE score (which was 1). To summarize, if the minimum DTW distance $\alpha$ exceeds our threshold $\beta$, we classify the entire pulse (and all its timesteps) as an artifact.

\subsubsection{GradCAM \& SHAP}
In this baseline experiment, we compare to the results achieved by Guo et al. \cite{sota}. Briefly, Guo et al. conducted their experiment by employing the Resnet-34 architecture proposed by Dai et al (2016) for 1D binary classification (identifying `clean' or `artifact'). This architecture, traditionally used for image classification, was repurposed for time series analysis via GradCAM \cite{gradcam}. Considering the relatively small size of the training dataset, they employed transfer learning on a pre-trained Resnet34-1D PPG signal quality classifier, as described by Zhang et al (2021) \cite{zhang}. This pre-trained model was originally trained on a UCSF PPG dataset acquired by Zhang et al \cite{zhang} from 3764 intensive-care unit patients according to Pereira et al (2019) \cite{pereira2019deep}. The last two residual blocks, global average pooling, and the final dense layer were subsequently retrained by Guo et al. To train a classification model, they created ground truth labels such that any signal that contained artifact timesteps was labeled as an artifact, while all other signals were labeled as non-artifacts. After this labeling process, the training set consisted of 175 clean signals and 3261 artifact signals.

The model was trained using the Adam optimizer \cite{kingma}, using the binary cross-entropy loss. An initial learning rate of $10^{-5}$ was set, which was scheduled to reduce to $5 \times 10^{-6}$ after 10 epochs, and further decrease to $1 \times 10^{-6}$ after 50 epochs. The maximum number of training epochs was capped at 100.

While the Resnet34-1D network is fundamentally a classifier, the goal is instead to obtain segmentation labels. Assuming that the model would focus on artifacts when tasked with predicting an artifact signal, Guo et al. employed two post-hoc explanatory methods to generate segmentation masks: SHAP \cite{shap} and GradCAM \cite{gradcam}.

SHAP is a unified measure of feature importance that assigns each feature an importance value for a particular prediction. It is based on game theory and computes Shapley values. The Shapley value of a feature represents the average marginal contribution of that feature across all possible feature subsets.

GradCAM utilizes the gradients of any target concept flowing into the final convolutional layer of a CNN to produce a coarse localization map highlighting the important regions in the image (or in this case, a time series) for predicting the concept. This is done by first computing the gradient of the output category with respect to feature maps of the last convolutional layer, then applying a global average pooling over the gradients to obtain weights for the feature maps. These weights are multiplied with the feature maps and summed over all maps to generate the final output. This output can be used as a heatmap, indicating which parts of the input were important in making the model's prediction. GradCAM can be applied to any CNN-based network without requiring architectural changes or re-training.

For the SHAP values, the computed values were first smoothed with a Gaussian filter. A binary segmentation mask was then created by choosing the class with the higher SHAP value at each timestep.

\subsubsection{Segade}
We compare against a high-performing and high-compute deep learning architecture, Segmentation-based Artifact Detection (Segade) by Guo et al. \cite{sota}. Segade is a 1D segmentation network similar in architecture to UNet, with a few changes; notably, the convolutions are replaced with residual convolutions for better gradient propagation and the convolutions are one-dimensional as the data type is one-dimensional. We refer to \cite{sota} for a detailed discussion of the architecture.

\subsubsection{Tiny-PPG}
Finally, we compare against the current state-of-the-art, Tiny-PPG \cite{yalizheng}, a light-weight deep convolutional neural network comprised of depth-wise separable convolutions and atrous spatial pyramid pooling trained via supervised cross-entropy loss and a contrastive loss. The original Tiny-PPG model was trained on the same train set as our method (DaLiA) but was not benchmarked against WESAD or TROIKA. Thus, we used their publicly available code to retrain 10 identical models from different starting initializations to establish test-set performance data on these datasets, while simultaneously verifying that we achieve nearly identical results on the DaLiA test set as the original paper reported. We leave the training details to the original work \cite{yalizheng}.

\subsubsection{ResNet-16}
We make slight modifications to an open-source 1D-ResNet model that was used to achieve competitive performance on 12-lead arrhythmia classification \cite{ribeiro_resnet} to accommodate our single-lead task. Specifically, we change the input channel width to 1 channel rather than 12, and slightly adjust strides and kernel sizes such that our data shapes remain coherent throughout the network. We train for 512 epochs over the dataset, with a batch size of 1024, with a learning rate of $0.001$, linearly decayed to $0.0$ over the course of training. We train with the AdamW optimizer in PyTorch ($\beta_1=0.9$, $\beta_2 = 0.999$, $\epsilon=10^{-8}$, weight decay $=10^{-2}$).

\subsubsection{Jia-CNN}
We use the same three convolutional neural network architectures as Jia et al. \cite{jiaCNN} (referred to as Jia-CNN), though we change the number of final output neurons from two to three to accommodate the difference in class number. Additionally, Jia-CNN was designed to receive an input signal at 250 Hz, thus we change only the resampling frequency of our preprocessing pipeline, leaving all else the same for an equal comparison. We follow the same training paradigm as ResNet-16, though we notably raise the learning rate to $0.005$ for Jia-CNN1 and Jia-CNN3. We set the learning rate to $0.002$ for Jia-CNN2, as we find that on some train splits, the model is unable to converge with higher learning rates.

\section{Results}
\subsection{PPG Segmentation}
\subsubsection{Test Set Performance}
We conduct our tests on several PPG datasets, DaLiA \cite{dalia}, WESAD \cite{wesad}, and TROIKA \cite{troika}, against several deep learning (Segade \cite{sota} and Tiny-PPG \cite{yalizheng}) and traditional machine learning methods. A full description of the datasets, pre-processing, and baseline experiments is given in the Methods section. We find that our largest model significantly exceeds the performance (measured in DICE score) of all baselines except the current state-of-the-art models (Segade and Tiny-PPG), for which it achieves $>$99\% of their DICE-score on DaLiA and WESAD, \textbf{with less than 2\% the parameter count of Segade and half the parameters of Tiny-PPG.} Our largest model even manages to \textbf{significantly outperform all other baselines on the most challenging and least in-distribution dataset (TROIKA)}. Our medium model, at 0.4\% of the parameter count of Segade, achieves at least 98\% of Segade's DICE score on all datasets and matches Segade's performance on TROIKA. Lastly, our smallest model, at 0.06\% of the parameter count of Segade, reaches 94\% of the DICE score of Segade on DaLiA and WESAD, though struggles on TROIKA, reaching 86\% of the DICE score of Segade. Detailed results and parameter counts are in Table \ref{tab:model_performance}. Furthermore, we find that our model is robust to noisy inputs, showing little degradation in DICE score when varying levels Gaussian noise are added to the test signal, even without augmenting training data with Gaussian noise (see Appendix Figure \ref{fig:noisy_signal}). We also find that it is empirically challenging to overfit on our train set even at high parameter counts and that there may be a potential performance ceiling due to data labeling inaccuracies (see Appendix \ref{sec:scaling_laws} and \ref{sec:appdx_data_lim}).

\begin{table*}[ht]
\centering
\begin{tabular}{|c|c|c|c|c|}
\hline
\textbf{Model} & \textbf{PPG-DaLiA} & \textbf{WESAD} & \textbf{TROIKA} & \textbf{Params} \\ \hline
Segade & $\mathit{0.8734 \pm 0.0018}$& $\mathit{0.9114 \pm 0.0033}$& $0.8050 \pm 0.0116$ & $2.3$M \\ \hline
Tiny-PPG & $\mathit{0.8781 \pm 0.0012}$& $\mathit{0.9141 \pm 0.0029}$& $\mathit{0.8246 \pm 0.0053}$& $85.9$K \\ \hline
Sliding window & $0.8068 \pm 0.0014$ & $0.8446 \pm 0.0013$ & $0.7247 \pm 0.0050$ & $51$K \\ \hline
Template matching & $0.6974 \pm 0.0323$ & $0.6954 \pm 0.0309$ & $0.6748 \pm 0.0122$ & N/A \\ \hline
ResNet-34 + GradCAM & $0.7129 \pm 0.0010$ & $0.7372 \pm 0.0024$ & $0.6989 \pm 0.0034$ & $4$M \\ \hline
ResNet-34 + SHAP & $0.6748 \pm 0.0000$ & $0.6634 \pm 0.0000$ & $0.6849 \pm 0.0001$ & $4$M \\ \hline
\textbf{SMoLK (Large)}& $\mathit{0.8692 \pm 0.0002}$ & $\mathit{0.9050 \pm 0.0002}$ & $\mathit{0.8322 \pm 0.0026}$ & $45.3$K \\ \hline
\textbf{SMoLK (Medium)}& $0.8610 \pm 0.0005$ & $0.8968 \pm 0.0008$ & $0.8067 \pm 0.0075$ & $8.5$K \\ \hline
\textbf{SMoLK (Small)}& $0.8191 \pm 0.0014$ & $0.8639 \pm 0.0015$ & $0.6891 \pm 0.0022$ & $1.4$K \\ \hline
\end{tabular}
\caption{\textbf{PPG Segmentation Performance.} Various baselines vs$.$ our approach at varying parameter counts on the PPG segmentation task, measured by DICE score. \textit{Italicized} results are at least 99\% of the performance of the best result. Our methods are in \textbf{bold}.}
\label{tab:model_performance}
\end{table*}

\subsubsection{Interpretability}
A considerable benefit of our approach is the clarity of the signal processing of the PPG, unlike in traditional deep neural networks. Since our model consists of a single layer of convolutions of varying kernel sizes, we can simply peer inside the model and determine how much each set of convolutions (and even each individual kernel) contributes to the overall classification signal.

To begin our exploration, we look to quantify the effect each kernel has on the output signal when presented with a perfect match to its signal. To compute the importance of the $m^{th}$ kernel, we calculate the following expression: 
\[\textit{kernel importance}_m = \left (\|\mathbf{k}_m\|_2^2  + b_m\right ) \cdot \textit{w}_m\]
where $\|\mathbf{k}_m\|_2^2 =\sum_jk^2_{m,j}$ is the sum of each squared component of the $m^{th}$ kernel, $\textit{b}_m$ is the $m^{th}$ kernel's bias, and $\textit{w}_m$ is the $m^{th}$ kernel's weight. This, by definition, computes the value added to the final signal by that kernel when the kernel perfectly overlaps with an exact template match. Furthermore, we also note that in the case of a normalized kernel and bias (i.e., $\sum_j\textbf{k}^2_{m,j} + b_m = 1$), the kernel importance directly equals the weighting of that kernel. We run this kernel analysis across all 10 cross-validation models for the small, medium, and large model sizes.

\begin{figure*}[h]
\centering
\includegraphics[width=1.0\textwidth]{"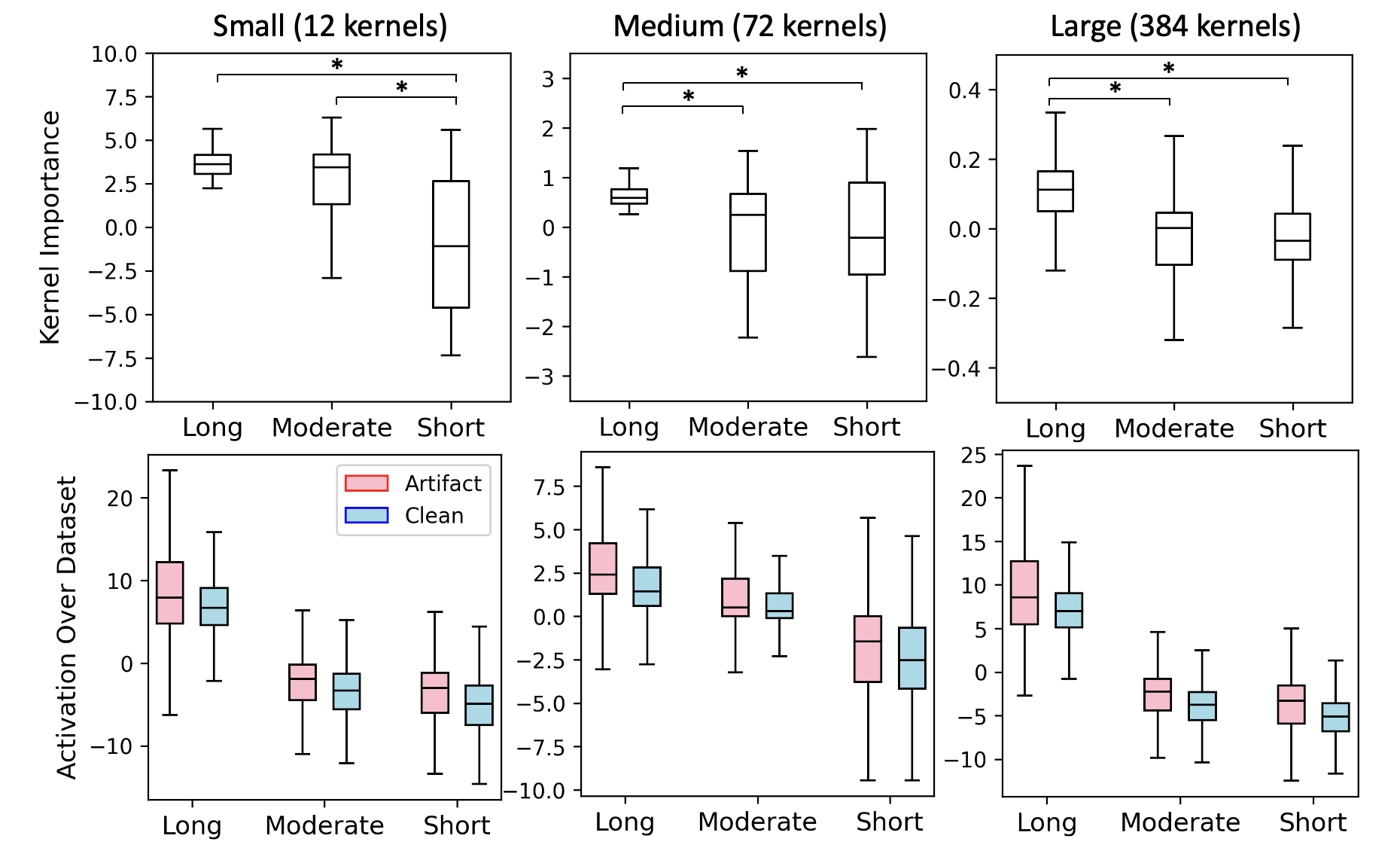"}
\caption{\textbf{Kernel Statistics.} The computed ``kernel importance'' of each kind of kernel in each model size (top row) and the empirical average activation of each kernel group observed over the clean and artifact parts of the dataset (bottom row). Significant differences in computed ``kernel importances'' are reflected in the empirical output values of those kernel groups over the test set. Significance was computed via a two-tailed independent samples t-test with Bonferroni correction. All differences are significant between the empirical kernel output due to the extremely large sample size. All box-and-whisker plots are constructed via the median value as the central line, the interquartile range (IQR) as the box, and the whiskers denoting the minimum and maximum value of the distribution. Outliers are defined as points that lie outside of $\pm 1.5\times \text{IQR}$ and were excluded from the plot for clarity, though all points were included in statistical analysis.}
\label{fig:kernel_effect}
\end{figure*}

We notice a clear pattern between kernel size and kernel importance. We find that for the medium and large models, the long kernels have significantly more positive ($p < 10^{-13}$) mean importance than the short and moderate kernels, with significance computed via two-tailed independent samples t-tests. For the small model, the long and moderate kernels have a mean importance significantly more positive than the small kernels ($p < 10^{-6}$) (see Figure \ref{fig:kernel_effect}). \textbf{These results indicate that long kernels generally learn to recognize poor-quality signal features, while medium and short kernels learn to recognize clean PPG features.}

We investigate this claim empirically by computing the activations of each kernel size group over the entire test dataset, grouping activation values by whether they were generated by artifact-containing PPG signals or clean signals. Essentially, we perform the convolution and weighting operation for each set of kernels grouped by size (long, moderate, or short), along the entire test dataset, concatenating the output values to one array. We further segregate the output values depending on whether they were generated by PPG signals labeled as artifact or clean. In the framework of our model, this is equivalent to computing:
\[
\sum_{m=1}^{M^g} \max(0, \mathbf{x} * \mathbf{k}^g_m + b_m^g) \cdot w_m^g
\]
where $g$ represents the kernel group in question, i.e., the long, moderate, or short kernels, $M^g$ is the number of kernels in group $g$, $\mathbf{b}^g$ are the biases of group $g$, and $\mathbf{w}^g$ are weights associated with kernel group $g$.

\begin{figure*}[h]
\centering
\includegraphics[width=12cm]{"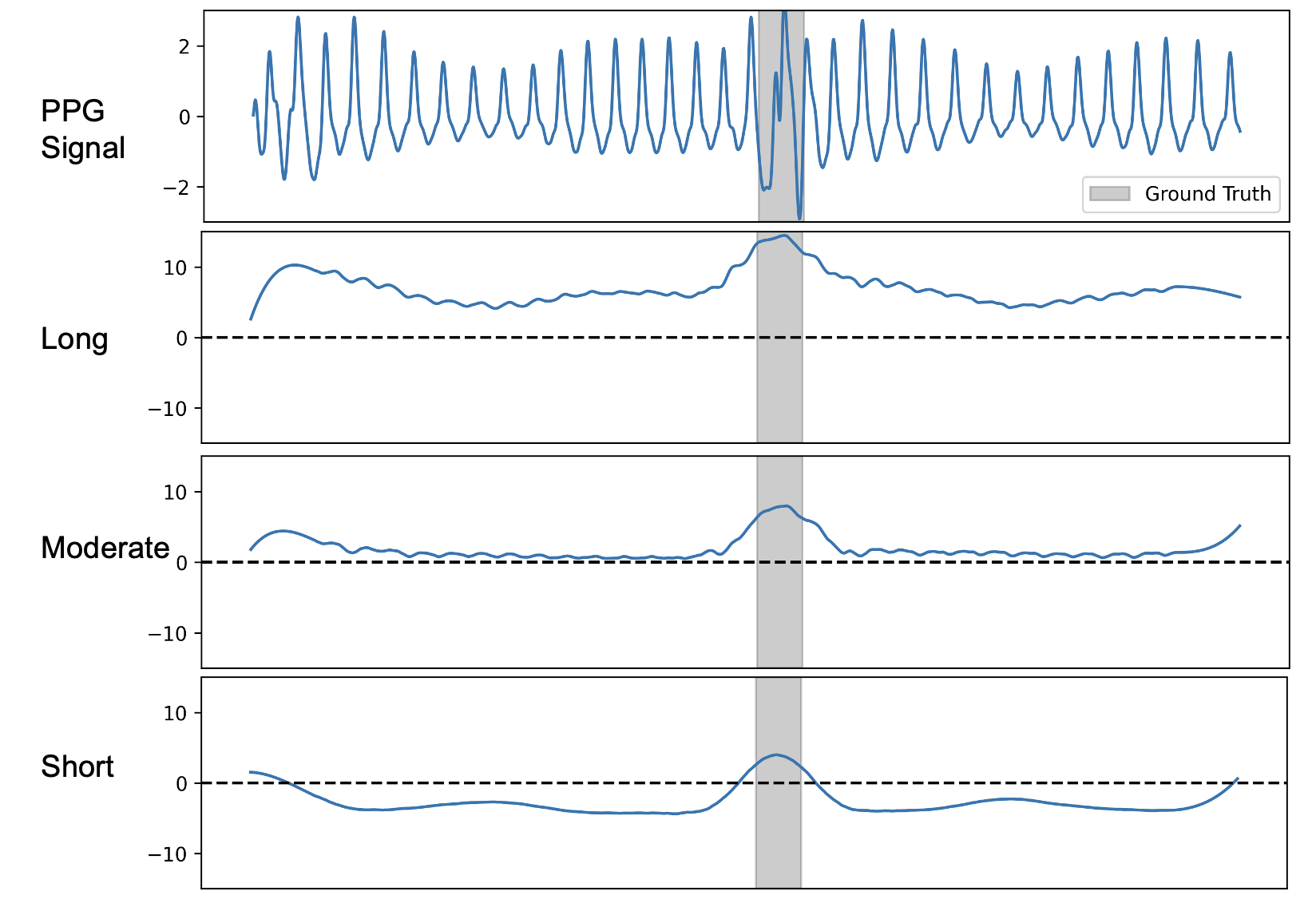"}
\caption{The contribution of each kernel group to the overall output signal. Noticeably, the ``long'' kernels have a positive signal even when convolved over clean segments, the ``moderate'' kernels have a signal close to zero when convolved over clean segments and a positive value over the artifact, and the ``small'' kernels have negative values except when convolved over the artifact.}
\label{fig:kernel_effect_example}
\end{figure*}

We find the claims generated by our mathematical analysis of kernel importance to be consistent with our empirical findings, with long kernels producing positive values most of the time, moderate kernels producing mostly negative or mixed positive/negative values (as seen in the medium-sized model), and short kernels producing mostly negative values, as shown in Figure \ref{fig:kernel_effect}. We demonstrate one such 30-second example of this in Figure \ref{fig:kernel_effect_example}. The agreement between our mathematical exploration of kernel importance and our empirical exploration of the values produced by various kernel groups over the test sets demonstrates that we can draw meaningful conclusions of our model without relying on post hoc explanability methods. These results are shown in Figure \ref{fig:kernel_effect} by comparing the top and bottom rows.

Next, we note that because our smallest model consists of only 12 kernels, we can easily visually inspect each kernel. They are all shown in Appendix Figure \ref{fig:12_kernel_model}. This is in stark contrast compared to most deep-learning models, where the parameters are typically so numerous that it is intractable to manually inspect each parameter; here, \textit{all} of the kernels can be shown on one page. Furthermore, parameters in deep neural networks are typically uninterpretable, whereas our model represents a template-matching paradigm for waveforms.

\subsubsection{Pruning Experiments}
We find that \textbf{applying our pruning procedure on long kernels provides the greatest parameter reduction with the least performance trade-off.} Pruning moderate and/or short kernels impacts performance more heavily and ultimately does not prune as many parameters as pruning long kernels, since each long kernel contains more parameters by definition. Ultimately, the choice of prune-to-performance ratio is up to the user and specific use case, though we provide example pruning schemes in Appendix Table \ref{tab:pruning} which attempt to maintain performance around 96\% of the mean original performance. Notably, as the starting model size increases, we are able to remove progressively more parameters with less of a decrease in performance. We hypothesize this may be due to increased capacity of larger models with greater learned redundancy. 

Furthermore, because our model is a well-behaved ``almost linear'' system, model weights can be na\"ively quantized with virtually no loss in performance. In fact, a simple cast from float32 to float16 yielded indistinguishable results. In contrast, the quantization process for a deep neural network is non-trivial, as error accumulates and propagates through the network.

\subsection{Atrial Fibrillation Detection}
\subsubsection{Classification Performance}
We tested our 12-kernel, 72-kernel, and 384-kernel models against a 16-layer ResNet and a previously state-of-the-art convolutional neural network. We train and validate on the computing in cardiology atrial fibrillation detection dataset \cite{cinc} and test on a holdout set collected by Zheng et al. \cite{Zheng2020, Zheng2022} (training/testing methodology and dataset descriptions are discussed in the Methods section). On the 10-fold cross-validation set, the medium and large-sized SMoLK models outperform or match all other baselines in terms of AUC-ROC, a critical clinical metric (Table \ref{table:learned-kernels-performance-train}), despite having dramatically fewer parameters (less than 1\% of ResNet-16's parameter count). On the hold-out test set, all SMoLK models achieve superior AUC-ROC to the other baselines for normal and atrial fibrillation detection (Table \ref{table:learned-kernels-performance-test}), though they underperform in detection of unknown ``other'' rhythms compared to the two highest parameter count models we tested: a 16-layer ResNet (referred to as ResNet-16) and a previously state-of-the-art convolutional neural network (referred to as Jia-CNN1).

\begin{table*}[ht]
\centering
\begin{tabular}{|c|c|c|c|c|c|}
\hline
\textbf{Model} & \textbf{F1} & \multicolumn{3}{c|}{\textbf{AUC-ROC}} & \textbf{Params} \\ \cline{3-5}
 & & \textbf{Normal} & \textbf{AFib} & \textbf{Other} & \\ \hline
\textbf{SMoLK (Large)} & $0.725 \pm 0.013$ & $0.873 \pm 0.009$ & $0.961 \pm 0.004$ & $0.809 \pm 0.015$ & $47.6$K \\ \hline
\textbf{SMoLK (Medium)} & $0.703 \pm 0.012$ & $0.864 \pm 0.010$ & $0.959 \pm 0.005$ & $0.790 \pm 0.013$ & $9.7$K \\ \hline
\textbf{SMoLK (Small)} & $0.637 \pm 0.015$ & $0.834 \pm 0.008$ & $0.934 \pm 0.008$ & $0.728 \pm 0.014$ & $2.4$K \\ \hline
ResNet-16 & $0.741 \pm 0.011$ & $0.852 \pm 0.008$ & $0.891 \pm 0.012$ & $0.799 \pm 0.012$ & $6.40$M \\ \hline
Jia-CNN1 & $0.710 \pm 0.029$ & $0.856 \pm 0.013$ & $0.900 \pm 0.026$ & $0.789 \pm 0.013$ & $93.6$K \\ \hline
Jia-CNN2 & $0.671 \pm 0.012$ & $0.838 \pm 0.010$ & $0.914 \pm 0.012$ & $0.749 \pm 0.014$ & $28.3$K \\ \hline
Jia-CNN3 & $0.720 \pm 0.011$ & $0.867 \pm 0.008$ & $0.934 \pm 0.012$ & $0.786 \pm 0.011$ & $33.4$K \\ \hline
\end{tabular}
\caption{\textbf{Atrial Fibrillation Detection.} Metrics on the train set, computed via 10-fold cross-validation.}
\label{table:learned-kernels-performance-train}
\end{table*}

\begin{table*}[ht]
\centering
\begin{tabular}{|c|c|c|c|c|c|}
\hline
\textbf{Model} & \textbf{F1} & \multicolumn{3}{c|}{\textbf{AUC-ROC}} & \textbf{Params} \\ \cline{3-5}
 & & \textbf{Normal} & \textbf{AFib} & \textbf{Other} & \\ \hline
\textbf{SMoLK (Large)}& $0.555 \pm 0.006$ & $0.953 \pm 0.003$ & $0.802 \pm 0.002$ & $0.551 \pm 0.011$ & \textcolor{blue}{$47.6$K}\\ \hline
\textbf{SMoLK (Medium)}& $0.549 \pm 0.008$ & $0.955 \pm 0.002$ & $0.798 \pm 0.004$ & $0.541 \pm 0.010$ & \textcolor{blue}{$9.7$K}\\ \hline
\textbf{SMoLK (Small)}& $0.538 \pm 0.010$ & $0.957 \pm 0.003$ & $0.780 \pm 0.008$ & $0.557 \pm 0.026$ & \textcolor{blue}{$2.4$K}\\ \hline
 ResNet-16& $0.567 \pm 0.008$ & $0.926 \pm 0.003$& $0.761 \pm 0.007$& $0.710 \pm 0.014$&\textcolor{red}{$6.40$M}\\ \hline
\end{tabular}
\caption{Metrics on test set, computed from each model trained on the 10-folds from the train set.}
\label{table:learned-kernels-performance-test}
\end{table*}

We also find that our models train and generalize significantly better in the low-data regime. Training on only 1\% of the dataset, our method achieves higher F1 and AUC-ROC scores than all baselines on both the train and holdout set, with ResNet-16 yielding performance consistent with random guessing. We hypothesized this could be due to overparameterization of our ResNet-16 model, thus as a control, we train a simple four-layer convolutional neural network with nearly the same parameter count as our large learned kernel model. We find that even when controlling for parameter count, deep neural networks struggle to learn in the low data regime, achieving near-random performance, while our method maintains a stable performance across parameter counts (Table \ref{table:low-data-test}). This is also evident at the varying parameter scales of the Jia-CNN models, though notably, Jia-CNN3 outperforms ResNet-16 in the low-data regime.

\begin{table*}[ht]
\centering
\begin{tabular}{|c|c|c|c|c|c|}
\hline
\textbf{Model} & \textbf{F1} & \multicolumn{3}{c|}{\textbf{AUC-ROC}} & \textbf{Params} \\ \cline{3-5}
 & & \textbf{Normal} & \textbf{AFib} & \textbf{Other} & \\ \hline
\textbf{SMoLK (Large)}& $0.410 \pm 0.017$ & $0.689 \pm 0.031$ & $0.589 \pm 0.020$ & $0.598 \pm 0.039$ & \textcolor{blue}{$47.6$K}\\ \hline
\textbf{SMoLK (Medium)}& $0.416 \pm 0.017$ & $0.699 \pm 0.023$ & $0.585 \pm 0.018$ & $0.606 \pm 0.037$ & \textcolor{blue}{$9.7$K}\\ \hline
\textbf{SMoLK (Small)}& $0.400 \pm 0.018$ & $0.674 \pm 0.023$ & $0.578 \pm 0.018$ & $0.591 \pm 0.034$ & \textcolor{blue}{$2.4$K}\\ \hline
ResNet-16 & $0.308 \pm 0.016$ & $0.512 \pm 0.017$ & $0.496 \pm 0.014$ & $0.512 \pm 0.012$ & \textcolor{red}{$6.40$M}\\ \hline
Small-CNN& $0.353 \pm 0.029$ & $0.628 \pm 0.041$ & $0.530 \pm 0.018$ & $0.603 \pm 0.034$ & \textcolor{blue}{$45.3$K} \\ \hline
\end{tabular}
\caption{Metrics on test set, computed from each model trained on the 10-folds from only 1\% of the data in the train set.}
\label{table:low-data-test}
\end{table*}

\subsubsection{Interpretability}
\begin{figure*}[h]
\centering
\includegraphics[width=12cm]{"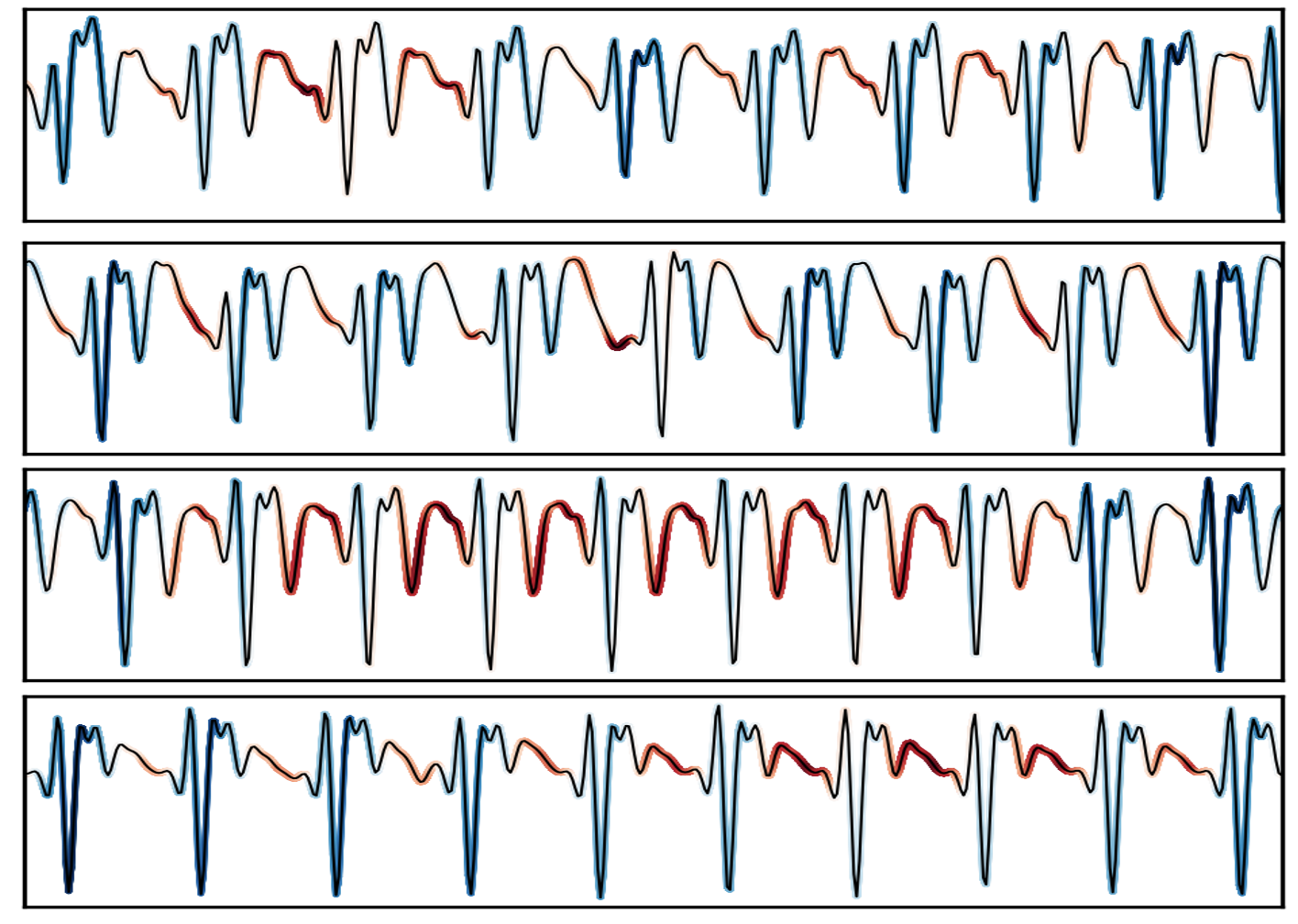"}
\caption{\textbf{Selected ECG Interpretations.} Various examples of $1^\circ$  atrioventricular block from the Zheng et al. dataset, along with the corresponding contribution maps as computed by Algorithm \ref{alg:contribution}. Redder colors indicate greater contribution to the $1^\circ$  atrioventricular block, while bluer colors indicate lesser contribution.}
\label{fig:ecg_interpretable}
\end{figure*}

To demonstrate the built-in interpretability of our method for classification tasks, we train a large learned kernel model to distinguish between $1^\circ$  atrioventricular block and sinus rhythm on the Zheng et al. dataset. We then apply Algorithm \ref{alg:contribution} to create a contribution map for each class. We find that our importance maps reliably identify the PR-interval (the defining factor) as the most contributory factor to the classification of $1^\circ$  atrioventricular block, despite never explicitly being given this signal (Figure \ref{fig:ecg_interpretable}). Importantly, this is not a post-hoc methodology (such as SHAP or GradCAM) that seeks to explain a model's behavior but is not guaranteed to be faithful. Instead, this method numerically computes the precise contribution of each part of the signal to the classification.

\section{Discussion}
We proposed a lightweight architecture for medical signal processing, which maintains performance on par or exceeding deep learning methods, all while remaining interpretable and efficient. We demonstrated the efficacy of our approach using PPG artifact detection and single-lead ECG atrial fibrillation detection as case studies. For PPG artifact detection, our approach achieved near state-of-the-art performance on three benchmark datasets with several orders of magnitude fewer parameters than deep learning methods and showed better generalization to out-of-distribution data, as evident by the markedly improved performance in the most out-of-distribution task, the TROIKA dataset. SMoLK's simplicity and interpretability allow for the inspection of the model's inner workings, providing insights into the contribution of each kernel to the overall prediction. Our method demonstrates extremely low compute and memory demands, setting it apart from deep-learning-based approaches. This enables high-quality signal assessment on low-power wearable devices.

Our results demonstrated that SMoLK effectively captures relevant features for identifying good and poor-quality PPG signals, and provides a directly interpretable, lightweight, and performant classifier. Our analysis of kernel importance revealed that larger kernels primarily contribute to the identification of poor-quality signal features, while smaller and medium-sized kernels capture normal PPG features. We demonstrate that our single-lead ECG classifier is directly interpretable, correctly identifying defining features of arrhythmias.
Importantly, our work demonstrates that for some complex tasks, simple and interpretable models can perform just as well as deep neural networks and are even preferable, in agreement with studies in other domains \cite{Rudin19}.

We developed a parameter reduction techniques to improve the compute-to-performance ratio, keeping in mind the low-power limitations of wearable devices. We demonstrated that we can remove over 22\% of the parameters in large models and maintain above 98\% of the original performance on a PPG segmentation task (Appendix Table \ref{tab:pruning}). Owing to the simplicity and near linearity of our model, we can na\"ively quantize the model weights with no impact on performance. This is in contrast to deep neural networks, where the extreme sensitivity of the system can amplify and propagate error due to quantization, leading to performance degradation. Quantization is also often a non-trivial process in deep neural networks as a result of dynamic range issues, again owing to the sensitivity of neural networks to small changes in parameter values. SMoLK suffers from none of the aforementioned complexities.

Our largest model can fit \textbf{in under 100 KB of memory when cast to float16} and can run in under 75 KB after weight pruning. \textbf{Our smallest model runs using under 3 KB of memory cast to float16}. Computing a raw (i.e., without post-process smoothing) signal score is extremely computationally efficient, with our largest model requiring on the order of 10 MFlops/second of signal processing. This memory-efficient and compute-efficient approach means that designers of low-power, wearable devices can now achieve state-of-the-art PPG signal quality assessment on extremely stringent compute budgets. SMoLK is so compute-efficient that it could feasibly be implemented on microcontrollers, such as consumer pulse oximeters.Additionally, implementation of our architecture is simple and straightforward to adapt to lower-level programming languages, whereas deep learning-based methods require the time-intensive re-implementation of complex operations into lower-level devices. Furthermore, for segmentation tasks, SMoLK's output only has local dependencies on the input signal; one does not need to wait for chunks of signal to be acquired before running the segmentation algorithm, nor does one need to stitch together multiple prediction windows. Instead, the output of the segmentation algorithm can be computed in real-time as the signal is acquired.

We also highlighted potential limitations in the datasets used for evaluation, as some labeling inaccuracies and discrepancies were observed in the test set samples with the poorest performance. This suggests that the practical maximum achievable performance for PPG artifact detection on these test sets may be limited by the quality of the available ground truth labels. Additionally, while we employed several hold-out sets for the PPG task to demonstrate the generalizability of our method, the datasets suffer from a small number of test subjects, limiting the statistical power of our result; further research is imperative to verify that the success of this methodology holds with a more diverse population. We evaluated our model on real-world noise, employing datasets whose distributions likely model the most common sources of noise (e.g., motion artifacts in the TROIKA dataset), though admittedly it was not possible to assess the robustness of our methodology to all possible noise sources. With all of this in mind, our methodology, and the field as a whole, would benefit from validation on large datasets composed of a large number of subjects under conditions that would generate a diversity of noise distributions. We have previously advocated for a broader effort \cite{RudinOpEd} to tackle this daunting task in future work. 

In conclusion, our method provides a robust, interpretable, and efficient approach for medical signal processing, making it competitive with deep learning approaches in performance and yet suitable for deployment on low-power devices. 

While our findings are promising, further research is needed to validate our approach in other medical signal-processing applications. The application of our method to other physiological signals and tasks could be explored, as well as the integration of our approach with other wearable devices like consumer pulse-oximeters, smart-rings, smart watches, and other low-profile devices for real-time signal quality assessment. This could significantly advance the field of wearable health technology, enabling continuous, real-time monitoring of biomarkers with high accuracy and minimal computational resources.

\section{Data Availability}
The data has been made publicly available at \hyperlink{https://doi.org/10.5281/zenodo.13117608}{https://doi.org/10.5281/zenodo.13117608} \cite{chen2024smolk}. We refer to the original works \cite{dalia, wesad, troika, cinc, Zheng2022, Zheng2020} for the unprocessed data, as well as maintain a copy of the pre-processed data in our GitHub repository \cite{chen2024smolk} (\hyperlink{https://github.com/SullyChen/SMoLK}{https://github.com/SullyChen/SMoLK}. The Computing in Cardiology dataset and Zheng et al. dataset are available on PhysioNet.org (\hyperlink{https://physionet.org/content/challenge-2017/1.0.0/}{https://physionet.org/content/challenge-2017/1.0.0/} and \hyperlink{https://doi.org/10.13026/wgex-er52}{https://doi.org/10.13026/wgex-er52}, respsectively).

\section{Code Availability}
The code has been made publicly available at \hyperlink{https://doi.org/10.5281/zenodo.13117608}{https://doi.org/10.5281/zenodo.13117608}, our GitHub repository \cite{chen2024smolk}: \hyperlink{https://github.com/SullyChen/SMoLK}{https://github.com/SullyChen/SMoLK}.

\section{Acknowledgements}
The authors would like to thank Anish Karpurapu for helpful discussions surrounding the interpretation of electrocardiograms. This work was supported by the National Heart, Lung, and Blood Institute (NHLBI) of the National Institutes of Health under grant number R01HL166233 (XH). We gratefully acknowledge this support.

\section{Competing Interests Statement}
The authors declare no competing interests.

\section{Author Contributions Statement}
Each author's contributions were significant enough to warrant inclusion in the authorship list for this manuscript. Sully F. Chen was the primary author, designed the study, and conceived the model architecture. Zhicheng Guo assisted heavily in the writing and editing of the manuscript and, along with Cheng Ding, ran the other baselines included in the paper and prepared the data for training and testing. Xiao Hu and Cynthia Rudin were the principal investigators and provided invaluable mentorship and assisted in the editing and writing of the manuscript.

\clearpage

\appendix

\section{Post-Processing Filters}
\label{sec:appdx_postprocessing}
We apply a $3^{rd}$ order Savitzky-Golay filter to our model output to produce a result consistent with the structure of PPG artifacts. However, we note that, excluding ultra-low kernel numbers, our model performs similarly on the test sets with and without this post-processing filter. It is thus an optional hyperparameter to be chosen by the user (Table \ref{table:appdx_performance_comparison}). 

\begin{table*}[h]
\centering
\caption{Comparison of performance scores with and without Savgol filtering across different datasets and kernel sizes.}
\label{table:appdx_performance_comparison}
\begin{tabular}{|l|cc|cc|cc|}
\hline
\multirow{2}{*}{Kernel Size} & \multicolumn{2}{c|}{DaLiA} & \multicolumn{2}{c|}{TROIKA} & \multicolumn{2}{c|}{WESAD} \\
\cline{2-7}
 & Savgol & None & Savgol & None & Savgol & None\\
\hline
12-kernel & 0.8185 & 0.7733 & 0.6889 & 0.6779 & 0.8632 & 0.8185 \\
\hline
72-kernel & 0.8607 & 0.8483 & 0.8114 & 0.7810 & 0.8966 & 0.8833 \\
\hline
384-kernel & 0.8690 & 0.8634 & 0.8331 & 0.8135 & 0.9047 & 0.8976 \\
\hline
\end{tabular}
\end{table*}

\section{Scaling Laws}
\label{sec:scaling_laws}
We explore the scaling behavior of our approach by gradually increasing the number of kernels and evaluating the performance on the test set. Our main objective is to understand the relationship between the parameter count and test-set performance, which helps us identify the potential limits and robustness of our approach.

\begin{figure}[h]
\centering
\includegraphics[width=8cm]{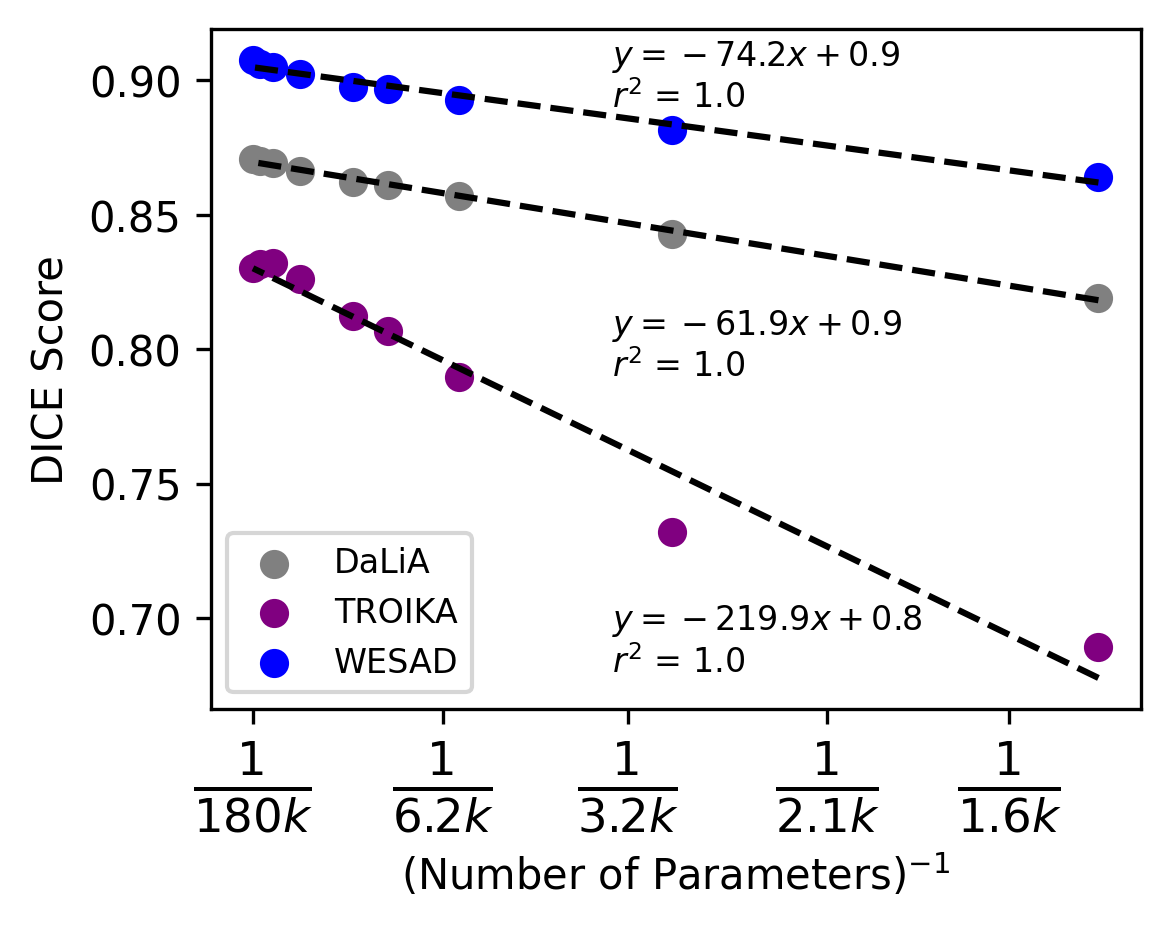}
\caption{Exploration of scaling laws reveals an asymptotic relationship between inverse parameter count and DICE score.}
\label{fig:asymp}
\end{figure}

To investigate the scaling behavior, we first conduct a series of experiments with different numbers of kernels and measure the corresponding test-set performance using the DICE score. As anticipated, we observe an asymptotic relationship between the parameter count and the test-set performance, as shown in Figure \ref{fig:asymp}.

Our analysis reveals that the test-set performance plateaus and remains roughly monotonically increasing as the parameter count increases, without exhibiting signs of overfitting. This finding suggests that \textbf{our approach is robust even at extremely high parameter counts.}

In summary, our exploration of scaling laws demonstrates an asymptotic relationship between parameter count and test-set performance, providing insights into the limits and robustness of our approach. The absence of overfitting at high parameter counts attests to the stability of our architecture.

We note that it is possible that a theoretical performance limit exists due to mislabeling in the test set; this is discussed further in the following section.

\section{Data Limitations}
\label{sec:appdx_data_lim}
The extrapolated performance ceiling for our approach prompted an inspection of the test set. We computed the accuracy of our approach on each 30-second sample in the test set, then sorted the test set by accuracy. Upon inspection of the test set samples on which our approach performed the poorest, we find that the source of this disagreement corresponds to suboptimal labeling rather than inaccurate classification by our approach. In other words, our method was correct and the labels were wrong.

\begin{figure*}[h]
\centering
\includegraphics[width=12cm]{"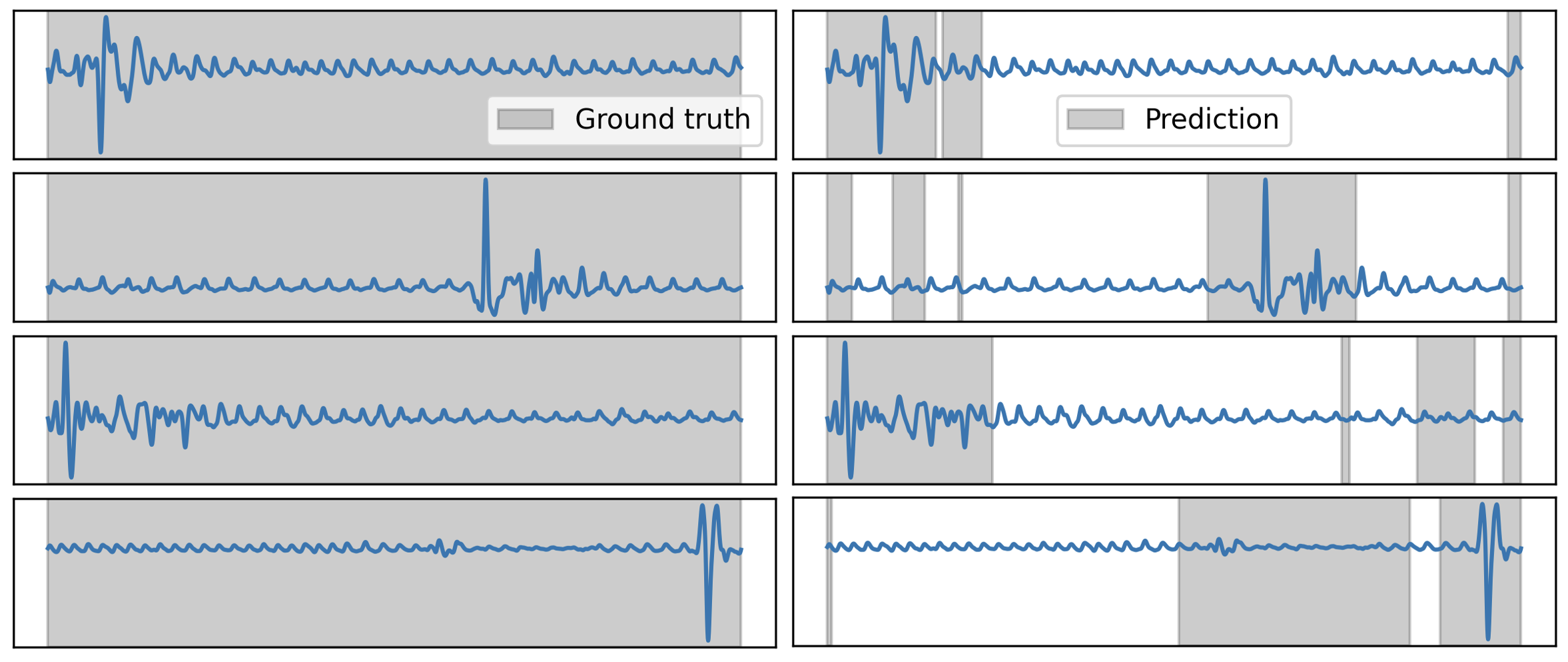"}
\caption{Ground truth labels for several 30-second PPG signal chunks are shown in comparison to their corresponding labels generated by SMoLK. Notably, low-amplitude PPG signals that still retain their waveform are marked as artifacts in the ground truth. These segments are not marked as artifacts by SMoLK, presumably because the signal retains a normal PPG waveform structure despite diminished amplitude, and because each 30-second PPG signal is normalized to its mean and standard deviation, thus eliminating amplitude changes locally.}
\label{fig:label_errors}
\end{figure*}

Several such inaccuracies are shown in Figure \ref{fig:label_errors}. Many of these inaccuracies in labeling seem to stem from PPG signals that have abnormally low amplitude compared to other signals in the dataset. Although this is unusual, we do not consider it an artifact, since the PPG signal retains its structure and signal-to-noise ratio. Furthermore, a PPG signal could have diminished amplitude due to factors like light travel distance, the brightness of the source light, lessened perfusion of the appendage (e.g., if the appendage is cold), etc.

\begin{figure*}[h]
\centering
\includegraphics[width=12cm]{"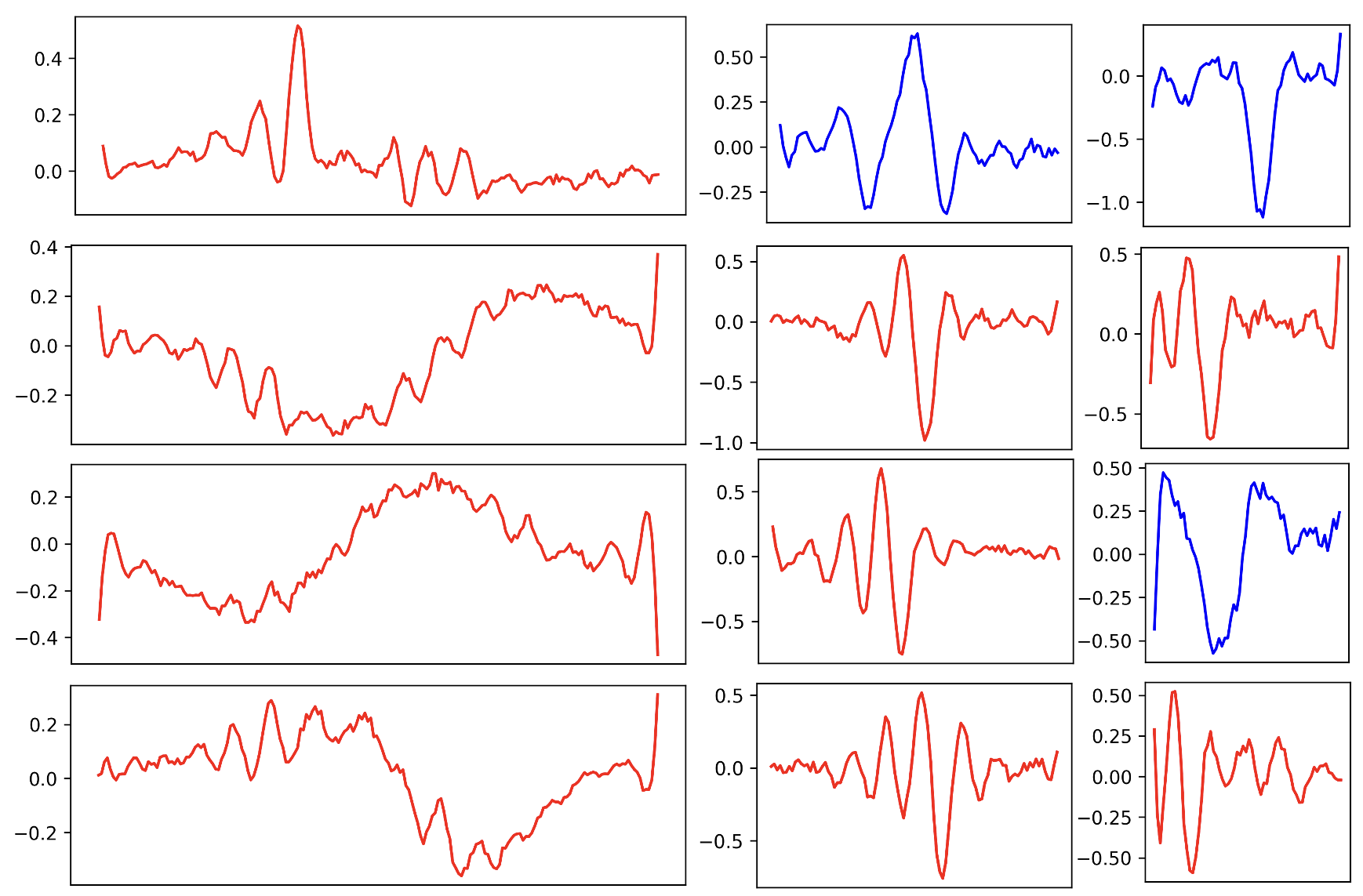"}
\caption{All 12 kernels of the small model. \textcolor{blue}{Blue} kernels contribute primarily to the detection of ``clean'' signals and \textcolor{red}{red} kernels contribute primarily to the detection of artifacts.}
\label{fig:12_kernel_model}
\end{figure*}

\section{Robustness to Noise}
\begin{figure*}[h]
\centering
\includegraphics[width=12cm]{"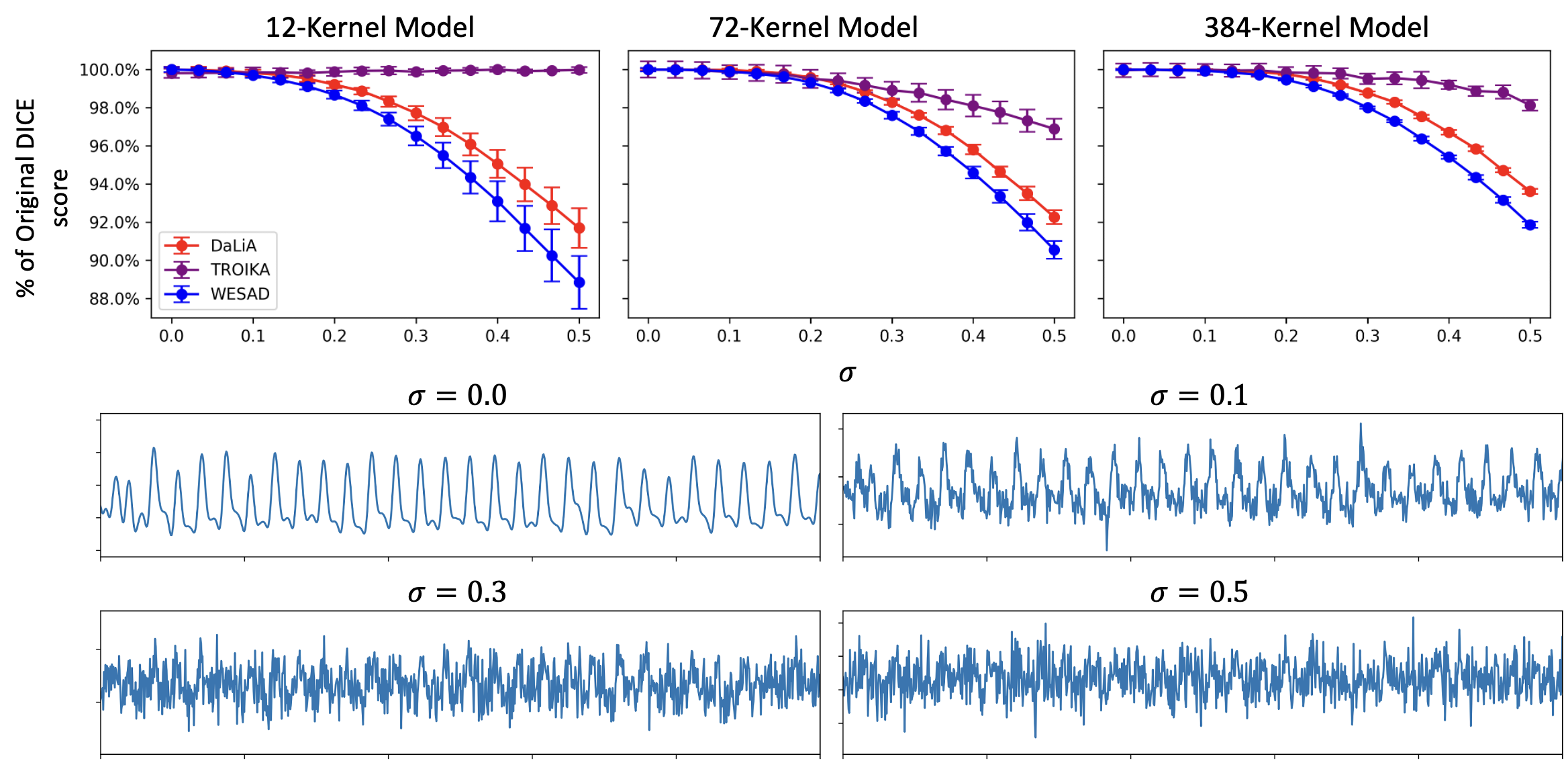"}
\caption{The performance of the learned kernel model as compared to baseline performance on the clean test signal (above panels). Examples of a PPG signals with various amounts of Gaussian noise added (bottom panels).}
\label{fig:noisy_signal}
\end{figure*}

To test our model's robustness to noise, we slowly add increasing amounts of Gaussian noise to our PPG segmentation test sets and evaluate the difference between the performance on the noisy signal and the original signal (Figure \ref{fig:noisy_signal}). We find that our learned kernel models are robust to large amounts of Gaussian noise, showing little degradation with a noticeably noisy signal.

\section{Pruning Results}
\label{sec:appdx_pruning}
We provide three example pruning results at various model scales, with a target performance decrease of no greater than 5\%. We would like to note that one can experiment with this pruning process via our open-source GitHub repository.

\begin{table*}[h]
\centering
\begin{tabular}{|c|c|c|c|c|c|}
\hline
\textbf{Model} & \textbf{Removed} & \textbf{All} & \textbf{DaLiA} & \textbf{TROIKA} & \textbf{WESAD} \\ \hline
Small & $16.4\% \pm 5.42\%$& $96.4\% \pm 4.6\%$& $94.8\% \pm 3.6\%$& $99.7\% \pm 0.8\%$& $94.5\% \pm 5.7\%$\\ \hline

Medium & $21.7\% \pm 2.5\%$& $94.9\% \pm 2.7\%$& $96.1\% \pm 1.5\%$& $92.7\% \pm 2.9\%$& $96.0\% \pm 2.1\%$\\ \hline

Large & $22.0\% \pm 1.8\%$& $98.1\% \pm 1.8\%$& $97.9\% \pm 1.5\%$& $98.8\% \pm 2.0\%$& $97.5\% \pm 1.7\%$\\ \hline
\end{tabular}
\caption{Pruning experiments conducted on various model sizes across all 10 folds, listing the percentage of parameters removed, the percentage of the original mean performance across all datasets, and the percentage of original performance on DaLiA, TROIKA, and WESAD.}
\label{tab:pruning}
\end{table*}

Furthermore, we show pruning results using various distance metrics perform relatively consistently, indicating a robustness to the chosen distance metric. We test three additional distance metrics on our Large model: cosine similarity between power spectra (cosine distance in frequency space), cosine similarity in feature space (the cosine similarity of the feature vectors themselves), and Manhattan distance.
\begin{table*}[h]
\centering
\begin{tabular}{|c|c|c|c|c|c|}
\hline
\textbf{Method} & \textbf{Removed} & \textbf{All} & \textbf{DaLiA} & \textbf{TROIKA} & \textbf{WESAD} \\ \hline
Power spectra & $19.1\% \pm 1.2\%$ & $96.4\% \pm 1.9\%$& $96.6\% \pm 1.4\%$& $97.2\% \pm 1.9\%$& $95.5\% \pm 1.9\%$\\ \hline

Cosine distance& $20.8\% \pm 1.4\%$& $95.7\% \pm 4.2\%$& $95.8\% \pm 3.9\%$ & $96.0\% \pm 
\% 2.7\%$& $95.3\% \pm 5.4\%$\\ \hline

Manhattan& $21.9\% \pm 1.9\%$& $98.3\% \pm 1.1\%$& $98.3\% \pm 0.9\%$& $98.6\% \pm 1.2\%$& $98.0\% \pm 1.3\%$\\ \hline
\end{tabular}
 \caption{Various additional distance metrics and their performance as compared to baseline.}
\label{tab:appdx_pruning_metrics}
\end{table*}

\end{document}